\documentclass[12pt]{article}

\usepackage{amssymb,amsfonts,eurosym,geometry,ulem,graphicx,caption,color,setspace,sectsty,comment,footmisc,caption,pdflscape,subfigure,array,import,bbm,booktabs,bookmark,longtable,titling,appendix,amsmath}

\usepackage{times}
\usepackage[round, authoryear]{natbib}

\newcolumntype{L}[1]{>{\raggedright\let\newline\\arraybackslash\hspace{0pt}}m{#1}}
\newcolumntype{C}[1]{>{\centering\let\newline\\arraybackslash\hspace{0pt}}m{#1}}
\newcolumntype{R}[1]{>{\raggedleft\let\newline\\arraybackslash\hspace{0pt}}m{#1}}

\title{\textbf{Buy Now, Pay Later (BNPL) \\...On Your Credit Card}\thanks{First version: January 5, 2022. The views expressed are the authors and do not necessarily reflect the views of the data provider. The data provider reviewed the paper before its release. We thank Michael Dowling \& Stefan Palan (the Editors), an anonymous reviewer, Anthony Lee Zhang, Constantine Yannelis, Karthik Srinivasan, Kilian Huber, Matthew Notowidigdo, Neale Mahoney, Pascal Noel, Scott Nelson, Walter Zhang, participants at Harvard Kennedy School, Federal Reserve Bank of Kansas City, NatWest Group roundtables, RAND Behavioral Finance Forum, Russell Sage Foundation Summer Institute in Behavioral Economics, Equifax Global Buy Now Pay Later (BNPL) Working Group, NBER Behavioral Public Economics Bootcamp, Financial Conduct Authority, \& industry participants for their comments and also to Arif Sulistiono and Fabian Gunzinger. This work is supported by the UK Economic and Social Research Council (ESRC) under grant number ES/V004867/1. `Real-time evaluation of the effects of COVID-19 and policy responses on consumer and small business finances'. Guttman-Kenney acknowledges support from the University of Chicago, Booth School of Business and the George J. Stigler Center for the Study of the Economy and the State.}}
 
\author{Benedict Guttman-Kenney\thanks{Corresponding author. bguttman@chicagobooth.edu University of Chicago, Booth School of Business, Chicago, IL, USA.} \and Chris Firth\thanks{christopher.firth@nottingham.ac.uk University of Nottingham, Department of Economics, Nottingham, UK.} \and John Gathergood\thanks{john.gathergood@nottingham.ac.uk University of Nottingham, Department of Economics, Nottingham, UK.}}

\date{\textit{Journal of Behavioral \& Experimental Finance, Forthcoming}}

\begin{document}

\maketitle

\begin{abstract}
\noindent We provide the first economic research on ‘buy now, pay later’ (BNPL): an unregulated FinTech credit product enabling consumers to defer payments into interest-free instalments.
We study BNPL using UK credit card transaction data.
We document consumers charging BNPL transactions to their credit card. 
Charging of BNPL to credit cards is most prevalent among younger consumers and those living in the most deprived geographies.
Charging a 0\% interest, amortizing BNPL debt to credit cards -- where typical interest rates are 20\% and amortization schedules decades-long -- raises doubts on these consumers' ability to pay for BNPL. This prompts a regulatory question as to whether consumers should be allowed to refinance their unsecured debt.
\end{abstract}



\section{Introduction}

`Buy now, pay later' (BNPL) is an unregulated FinTech credit product enabling consumers to defer payments interest-free into one or more (often four or fewer) instalments.
With \pounds2.7bn in UK BNPL lending during 2020, the UK BNPL market is larger by volume of lending than the UK payday loan market at its peak.
Rapid recent growth in consumer use of BNPL have led UK, EU, and US governments and consumer financial protection regulators to consider whether and how to regulate BNPL, while Australia recently introduced regulatory measures.\footnote{US \href{https://www.consumerfinance.gov/about-us/newsroom/consumer-financial-protection-bureau-opens-inquiry-into-buy-now-pay-later-credit/}{https://www.consumerfinance.gov/about-us/newsroom/consumer-financial-protection-bureau-opens-inquiry-into-buy-now-pay-later-credit/} \& \href{https://financialservices.house.gov/events/eventsingle.aspx?EventID=408594}{https://financialservices.house.gov/events/eventsingle.aspx?EventID=408594}.
UK: \href{https://www.fca.org.uk/news/press-releases/fca-publishes-woolard-review-unsecured-credit-market}{https://www.fca.org.uk/news/press-releases/fca-publishes-woolard-review-unsecured-credit-market} \&
\href{https://www.gov.uk/government/consultations/regulation-of-buy-now-pay-later-consultation}{https://www.gov.uk/government/consultations/regulation-of-buy-now-pay-later-consultation.}.
EU: \href{https://ec.europa.eu/info/sites/default/files/new\_proposal\_ccd\_en\_3.pdf}{https://ec.europa.eu/info/sites/default/files/new\_proposal\_ccd\_en\_3.pdf}.
Australia: \href{https://asic.gov.au/about-asic/news-centre/find-a-media-release/2020-releases/20-280mr-asic-releases-latest-data-on-buy-now-pay-later-industry/}{https://asic.gov.au/about-asic/news-centre/find-a-media-release/2020-releases/20-280mr-asic-releases-latest-data-on-buy-now-pay-later-industry/}. UK payday loans: \cite{gathergood2019payday}.}

Very little is known about how consumers use BNPL products, despite the market's rapid growth, and scrutiny by consumer financial protection regulators.
At the time of writing, in December 2021, there were no relevant economics or finance research papers studying BNPL.
Searches for `Buy Now, Pay Later' and `BNPL' on ArXiv returned no results, on SSRN returned a single law working paper on the Singapore BNPL market \citep{sng2021buy}, there were no relevant results on NBER Working Papers beyond \cite{berg2021fintech}'s broader review of FinTech lending, and on google scholar showed no economics or finance papers in either published or working form (the only work appearing was from other academic fields).\footnote{See Online Appendix for more details including consumer surveys, blog discussions, and industry reports.}

We present the first quantitative academic research to study consumer use of BNPL.
Given the lack of prior literature, we start our paper by providing institutional details on BNPL to help stimulate research on this topic.

One explanation for the lack of prior research is a lack of data: for example, most BNPL is not visible in credit files.
To help fill this gap, we offer early insights into how consumers use BNPL, by analysing BNPL transactions that show in UK credit card transaction data.\footnote{See \citealp{agarwal2009payday} for an example of early analysis of consumer use of payday loans informing subsequent causal research \citep[e.g.][]{gathergood2019payday,allcott2021high}}

BNPL is often presented as a standalone credit product in the form of an interest-free loan repaid in one or more instalments.
At the point of purchase, the consumer provides details of the account from which the payments will be taken, after which loan repayments occur. 
However, unlike other credit products, consumers can choose to repay BNPL using a credit card.

We research whether consumers charge BNPL transactions to their credit card, and the degree of heterogeneity in such behaviors across consumers.
We find the existence of UK consumers charging their interest-free BNPL to their credit card.
Such behavior is more common among (i) younger consumers and (ii) consumers living in more deprived geographies. 
Taken together these facts raise a concern that some consumers will enter into a debt spiral. 
A debt spiral may occur from transforming a 0\% interest BNPL debt that amortizes over a few weeks or months into credit card debt: a product that typically incurs 20\% interest rates and has
decades-long amortization schedules.
It prompts a regulatory question as to whether consumers should be allowed to refinance their unsecured debt.

\section{Institutional Details}

\subsection{BNPL Product Structures}

Buy Now, Pay Later (BNPL) is provided at the point of sale, providing consumers the option to defer payments into one or more interest-free instalments.\footnote{See Online Appendix for more institutional detail on BNPL.}
BNPL is mainly used online with lenders typically a third party separate to the retailer.
Most BNPL is unregulated.
We analyze the UK BNPL market but the US market is analogous.

Repayment structures vary across and within BNPL lenders.
For example, Klarna in the UK provides an option to repay in the next thirty days as well as an option to repay in three instalments thirty days apart whereas Clearpay has four payments that are two weeks apart. In the US, the Consumer Financial Protection Bureau defines a BNPL as having 4 or fewer instalments but some products have more (e.g. OpenPay in the UK offers 3-10 instalments).
While BNPL are primarily instalment loans, there is product innovation in structures.
For example, Amazon features a credit limit (`Instalments by Barclays') while some debit and credit cards enable cardholders to pay particular transactions in BNPL-esque instalments (e.g. American Express's `Pay It Plan It').

BNPL lenders primarily generate revenue through merchant fees of 3-6\% rather than charging interest or fees.\footnote{\href{https://www.consumerfinance.gov/about-us/newsroom/consumer-financial-protection-bureau-opens-inquiry-into-buy-now-pay-later-credit/}{https://www.consumerfinance.gov/about-us/newsroom/consumer-financial-protection-bureau-opens-inquiry-into-buy-now-pay-later-credit/}}
Not all BNPL lenders charge late fees - for example PayPal does not. 
Non-payment of BNPL debt can still have consequences: BNPL lenders may block new purchases by that consumer, pass unpaid debt to debt collectors, and missed payments may get reported in credit files.
Consumers may experience costs if their BNPL payment has knock-on adverse effects on their other finances such as triggering overdraft fees, accumulating credit card interest, missing payments on other bills (as found in the UK payday lending market by \citealp{gathergood2019payday}).

\subsection{Economics of BNPL}

Economic theory provides mixed perspectives on the consumer benefits of deferred payments through credit products such as BNPL.
The life-cycle model implies opportunities to smooth consumption at zero interest cost are weakly welfare improving -- especially in a high inflation environment \citep{ando1963life}.

Deferred payments -- through BNPL or other credit products -- also presents the possibility for welfare losses. 
For example, financially unsophisticated or na\"ive present focused consumers mistakenly overconsuming \citep[e.g.][]{allcott2021high}.

A key BNPL product feature is decoupling the consumption benefit today from the pain of paying later \citep[e.g.][]{prelec1998red} -- something especially attractive to present biased consumers \citep[e.g.][]{o1999doing}. 

\subsection{BNPL on Credit Cards}

Lenders providing regulated credit are required to consider, at the time of credit application, the ability of a consumer to repay a debt -- known as ability to pay in the US and affordability or creditworthiness in the UK -- out of their income or assets \textit{``without the customer having to borrow to meet the repayments''}.\footnote{\href{https://www.fca.org.uk/publication/policy/ps18-19.pdf}{https://www.fca.org.uk/publication/policy/ps18-19.pdf}. See \cite{guttman2017preventing} \& \cite{defusco2020regulating} for research into UK and US ability to pay rules.}
This means, for example, a consumer cannot make a payment for their regulated UK mortgage with their credit card (or meet one credit card payment with another credit card).
This applies irrespective of whether a consumer repays their credit card in full or revolves interest-bearing debt.
These consumer protection regulations are designed to protect vulnerable consumers from harm such as cases where the incentives of lenders are misaligned with the incentives to improve consumer outcomes.

As BNPL is unregulated it is not subject to such ability to pay regulations: BNPL can be charged to debit or credit cards.\footnote{In the UK, debit cards are the most common means of payment. 57\% all UK payments are made on debit or credit cards. Cash has declined to from 45\% to 15\% of payments 2015 to 2021. The remainder of transactions are mainly bank transfers (e.g. Autopays/Direct Debits). Source: UK Finance, UK Payments Markets Summary, August 2022.}
Transferring BNPL debt to credit cards increases the risks to the consumer and thus is a warning flag to regulators.
Credit cards can be a more costly form of credit with an average interest rate of near 20\% APRs unless the cardholder repays their balance in full.
Credit cards also have decades long amortization schedules if the cardholder only makes the minimum payment -- a costly behavior common in UK and US data \citep{keys2019minimum,adamssem}.
The burden of persistently carrying credit card debt may also have non-financial costs such as adverse mental health impacts.
As one BNPL lender stated ``there is clearly greater risk of consumer harm from spending on credit cards'' (than BNPL).\footnote{\href{https://www.cityam.com/buy-now-pay-later-firm-klarna-claims-credit-cards-pose-the-real-risk/}{https://www.cityam.com/buy-now-pay-later-firm-klarna-claims-credit-cards-pose-the-real-risk/}}
If a consumer repays their credit card in full and does not have binding liquidity constraints (i.e. not near their credit card limit and have liquid cash available) there is limited benefit from charging BNPL to their credit card.
This is because any BNPL payments that are charged within the same credit card payment cycle (with its interest-free grace period that lasts 15-56 days) all come due at the same time.
Any payments that come due in subsequent payment cycles effectively extend the grace period before payments are due.
However, this comes at a cost since it is also discounting the benefits of any credit card rewards points relative to if the consumer had immediately charged the transaction to their credit card in one instant (non-BNPL) payment.\footnote{There is also an effort cost of the consumer having another financial intermediary to interact with.}
Finally, if a consumer is unable to meet their BNPL payments on-time, transferring BNPL debt to their credit card may be simply postponing an inevitable default and risk shifting the credit losses from the BNPL lender to the credit card lender.
There is evidence of this: Capital One deemed the risks so great it has banned consumers from charging BNPL to their global credit cards.\footnote{\href{https://www.reuters.com/article/us-capital-one-fin-payments/capital-one-stops-risky-buy-now-pay-later-credit-card-transactions-idUSKBN28H0OR}{https://www.reuters.com/article/us-capital-one-fin-payments/capital-one-stops-risky-buy-now-pay-later-credit-card-transactions-idUSKBN28H0OR}}

\section{Data}

We use anonymized UK credit card transactions data sourced from multiple banks and credit card issuers.\footnote{See \cite{baker2021household} for a review of household financial transaction data.}
These data are created to be a real-time leading indicator used by industry to track spending (in aggregate and also of particular firms, sectors, and regions) and are provided to us under an academic licence.\footnote{These are similar to data used in the US by \cite{chetty2020real} to track the COVID-19 pandemic in real-time. See \cite{vavra2021tracking} for a review of using private sector administrative micro data for tracking the COVID-19 pandemic.}
Our data include approximately one million credit cards held by UK consumers between December 2018 \& September 2022.
For context, the median UK credit cardholder holds one credit card \citep{gathergood2016can,gathergood2019individuals}.
The Online Appendix contains additional descriptive analysis on these data.

Each transaction record provides details including the spending amount and tagged information on the type of spending.
Each transaction has an anonymized card-account identifier to enable tracking over time.
For each card-account we observe the cardholder's age range and geography (`postcode sector').
Postcode sectors are very granular geographies: there are over 11,000 postcode sectors in the UK with each sector containing approximately 3,000 addresses.
As these are a dataset of credit card spending they do not include when or whether the consumer made credit card repayments, their cost of borrowing, whether they made payments via debit cards, or have unpaid BNPL payments due. 

BNPL transactions are tagged by us via a field that records a transaction's payment processor (e.g. Afterpay, Klarna).
We do not categorize PayPal as a BNPL lender in our analysis as we cannot distinguish BNPL from its large, established non-BNPL business of processing payments.
Klarna are the largest BNPL lender in the UK but they also provide some payment processing services so we may be tagging some transactions processed by them that are not BNPL.
Other BNPL lenders -- most notably Clearpay -- are only BNPL lenders without payment processing services.\footnote{Square is a payment processing business that recently acquired Afterpay (known as Clearpay in the UK). As they have different names we can differentiate Afterpay/Clearpay's BNPL from Square's payment processing in our transactions data.} 

Although our data is a sample and subject to potential measurement error, we feel justified using it as an informative proxy for all BNPL usage in a market that currently has very thin data and, of course, as a highly accurate reading of BNPL usage on credit cards. 
The latter is especially important for
regulators wishing to put in place measures that help consumers avoid high costs of debt, or at the
extreme, entering debt spirals. 
We believe our data is an informative proxy given the absence of representative, comprehensive data: a matched, transaction-level dataset of BNPL and credit cards.

\section{Results}

\subsection{BNPL Use On Credit Cards}

Our first main finding is documenting the existence of charging of BNPL transactions to credit cards. 
BNPL transactions are commonly present on credit cards: 19.5\% of UK credit cards active (i.e. with any transaction present) in December 2021 have a transaction by a BNPL firm charged to their credit card during 2021.

Some cardholders have multiple BNPL transactions on their credit cards.
This may be due to multiple BNPL purchases or BNPL purchases split into multiple instalments.
While each individual BNPL transaction is typically small (a median value of \pounds19.65 and 96\% are \pounds100 or less), the total amount of BNPL transactions per card during 2021 by credit cardholders using BNPL is non-trivial: the median value is \pounds157 and 17.6\% have spent \pounds500 or more.\footnote{Understanding repeated use of BNPL is an important avenue for future research. 
For example, repeating analyses similar to that done on payday loans to understand consumer patterns of BNPL borrowing and how commonly they ultimately end up in default or a cycle of persistent debt. \href{https://files.consumerfinance.gov/f/201403\_cfpb\_report\_payday-lending.pdf}{https://files.consumerfinance.gov/f/201403\_cfpb\_report\_payday-lending.pdf}}

What fraction of all BNPL is charged to credit cards? 
In the US this is estimated to be 22\%.\footnote{\href{https://www.creditkarma.com/about/commentary/consumers-rely-on-buy-now-pay-later-amid-record-inflation-use-credit-to-pay-it-off}{https://www.creditkarma.com/about/commentary/consumers-rely-on-buy-now-pay-later-amid-record-inflation-use-credit-to-pay-it-off}}
For the UK, a consumer survey estimates 26\% of BNPL consumers paid for BNPL using their credit card and 42\% using any type of borrowing (e.g. credit card, overdraft, payday loan).\footnote{\href{https://www.citizensadvice.org.uk/about-us/about-us1/media/press-releases/two-fifths-borrowed-to-pay-off-buy-now-pay-later/}{https://www.citizensadvice.org.uk/about-us/about-us1/media/press-releases/two-fifths-borrowed-to-pay-off-buy-now-pay-later/}.}
The incidence of such practices may vary across BNPL lenders and over time.\footnote{One anonymous UK BNPL lender reported ``less than 10\% of its customers repay using a credit card''. \href{https://www.ukfinance.org.uk/system/files/UK-Finance-response-to-the-HMT-BNPL-consultation-FINAL-060122.pdf}{https://www.ukfinance.org.uk/system/files/UK-Finance-response-to-the-HMT-BNPL-consultation-FINAL-060122.pdf}}

The patterns of transactions we observe are consistent with a variety of potential economic models of consumer behavior.
One plausible model is where overconfident consumers mispredict their future ability to repay BNPL on-time and, when payments come due, defer payments by charging them to their credit card -- potentially transferring them interest-bearing credit card balances or defaulting on this debt.

\subsection{BNPL Usage Over Time}

\begin{figure}[!htb]
  \caption{Value of BNPL transactions on active credit cards (black line) and active debit cards (yellow line), 2019 - 2022}
    \label{fig:ts}
    \begin{center}
    \begin{tabular}{c} 
    {\includegraphics[width=5.25in]{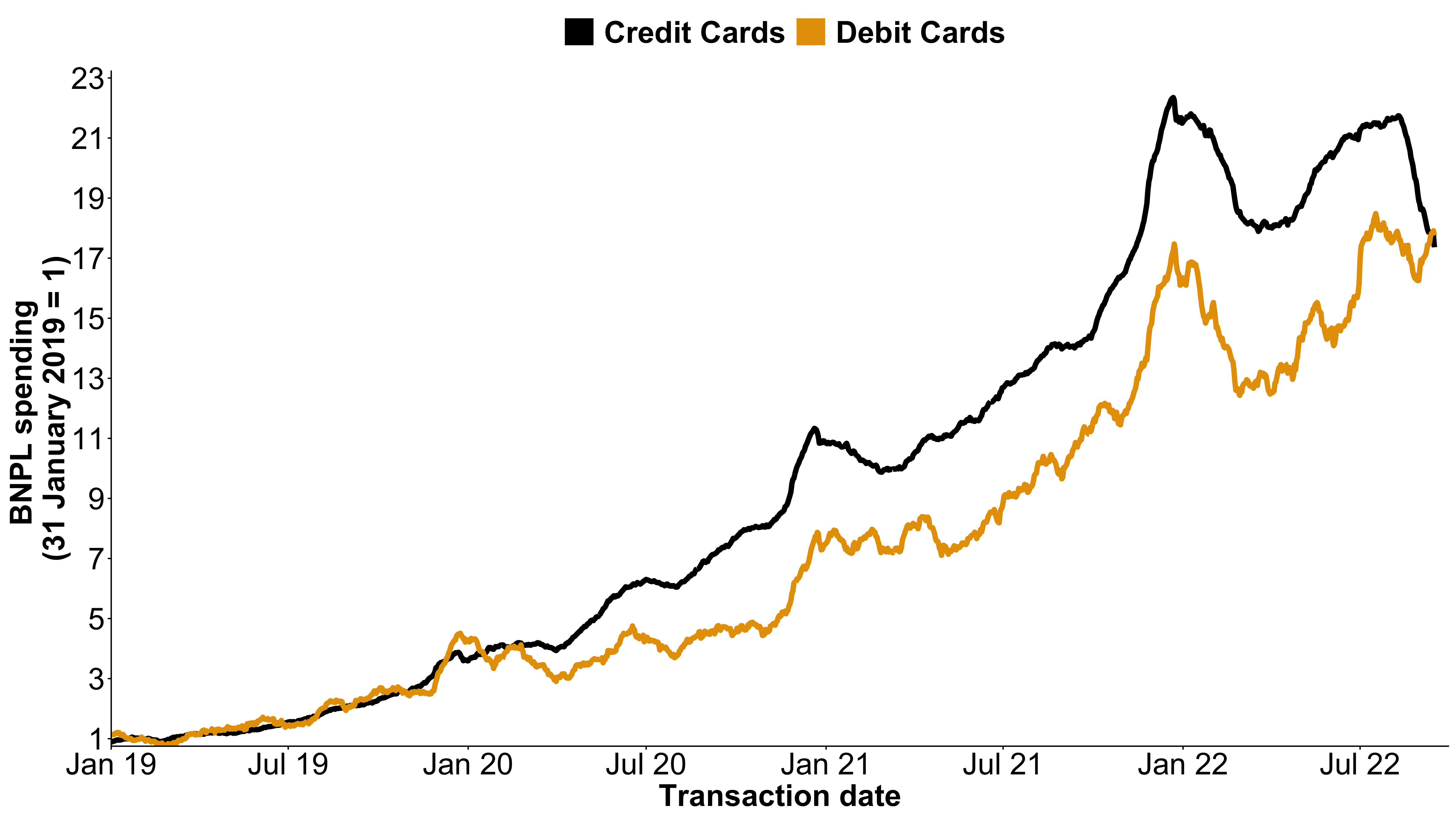}} 
    \end{tabular}
    \end{center}

    \begin{singlespace}
    \noindent {\footnotesize \textit{Notes: UK credit and debit card transactions. BNPL is buy now, pay later. Each series is 28 day moving averages of value of BNPL spending on those cards indexed to 1 in January 2019. 19.5\% of UK credit cards active (i.e. with any transaction present) in December 2021 have a transaction by a BNPL firm charged to their credit card during 2021, while 15.3\% and 12.1\% have charged in the last six and three months leading up to December 2021.}}
    \end{singlespace}
\end{figure}

Figure \ref{fig:ts} shows rapid growth in the value of BNPL spending on credit cards between January 2019 and December 2021: increasing 21.4 times.
To help interpret our time series of BNPL spending on credit cards, Figure \ref{fig:ts}  also includes an index of BNPL payments on debit cards -- using a separate sample from the same data provider. 
Both series follow similar trends.

BNPL spending on credit cards as a percent of all credit card spending averaged 1.2\% and 1.6\% during 2021 and December 2021.
This rapid growth lines up with official estimates which size the value of all UK BNPL lending during 2020 at \pounds2.7bn - ``more than tripling in 2020'': in our data the 2020 value of BNPL transactions is 3.4 times its 2019 levels.\footnote{\href{https://www.fca.org.uk/news/press-releases/fca-publishes-woolard-review-unsecured-credit-market}{https://www.fca.org.uk/news/press-releases/fca-publishes-woolard-review-unsecured-credit-market} \&
\href{https://www.gov.uk/government/consultations/regulation-of-buy-now-pay-later-consultation}{https://www.gov.uk/government/consultations/regulation-of-buy-now-pay-later-consultation.}}

December 2021 is the peak in BNPL spending on credit cards by transactions value and the index is 5.9 times December 2019 and 2.0 times its December 2020 levels.
There are local seasonal peaks aligning to the timing of payments coming due for `Black Friday' and Christmas spending.
BNPL on credit cards grew during the first wave of COVID-19 (March - April 2020).
These were a period when UK consumption sharply fell. 
BNPL on credit cards growing during this period may be a function of consumers shifting towards spending online during this period.
Our real-time data also reveals a 2022 slump in BNPL on credit cards following worsening macroeconomic conditions: a period when BNPL lenders' valuations also fell.\footnote{\href{https://www.ft.com/content/483451db-9221-4ca4-83a6-b4ddc6bfcfbb}{https://www.ft.com/content/483451db-9221-4ca4-83a6-b4ddc6bfcfbb}}

\subsection{BNPL Usage By Age}

\begin{figure}[!htb]
  \caption{BNPL usage on active credit cards by age, 2021}
    \label{fig:age}
    \begin{center}
    \begin{tabular}{c} 
    \textbf{A. BNPL usage (\% credit cards)}  \\ 
    {\includegraphics[width=6in]{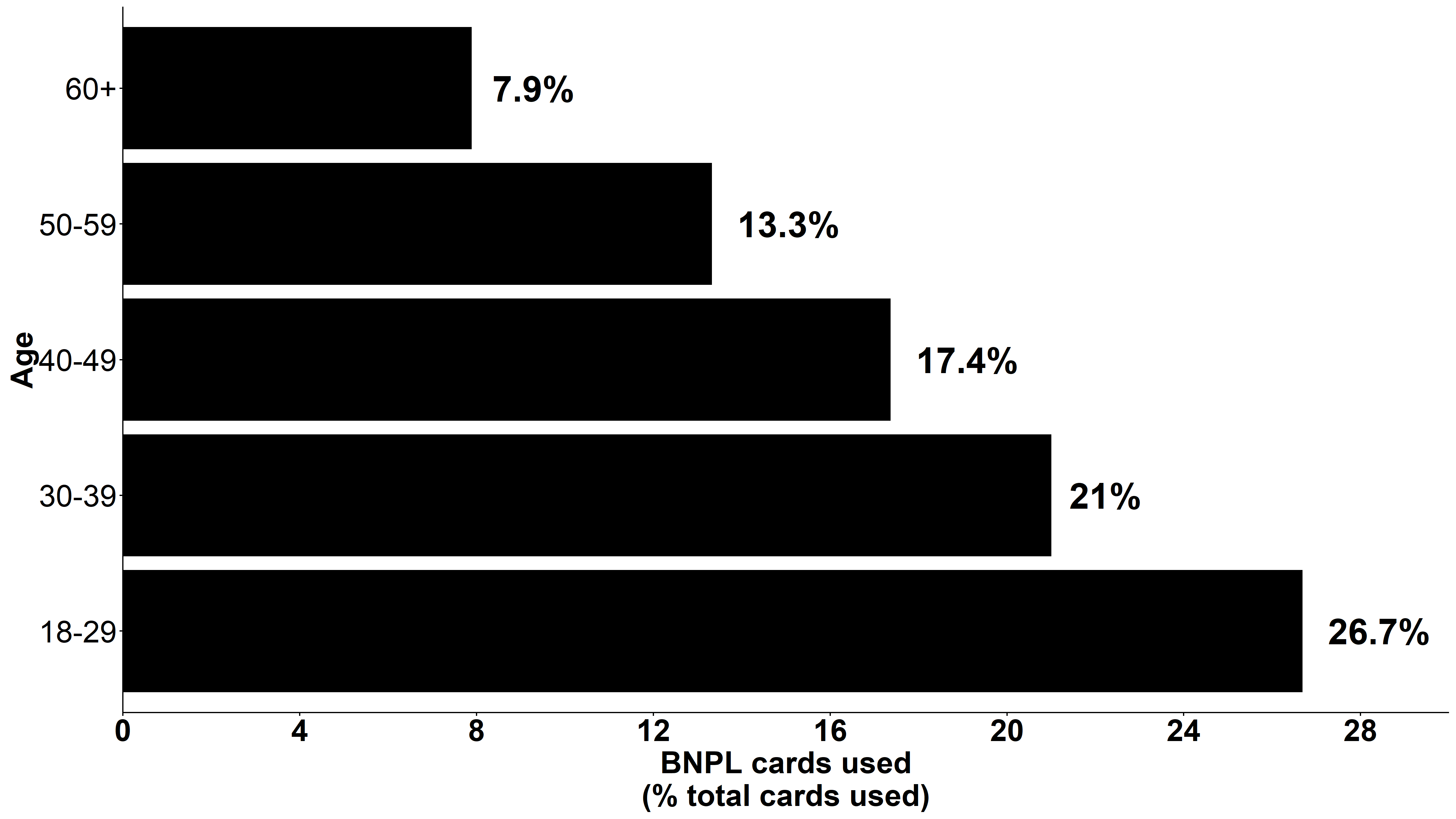}} \\ \\ 
    \textbf{B. BNPL usage (\% credit card spending)} \\ 
    {\includegraphics[width=6in]{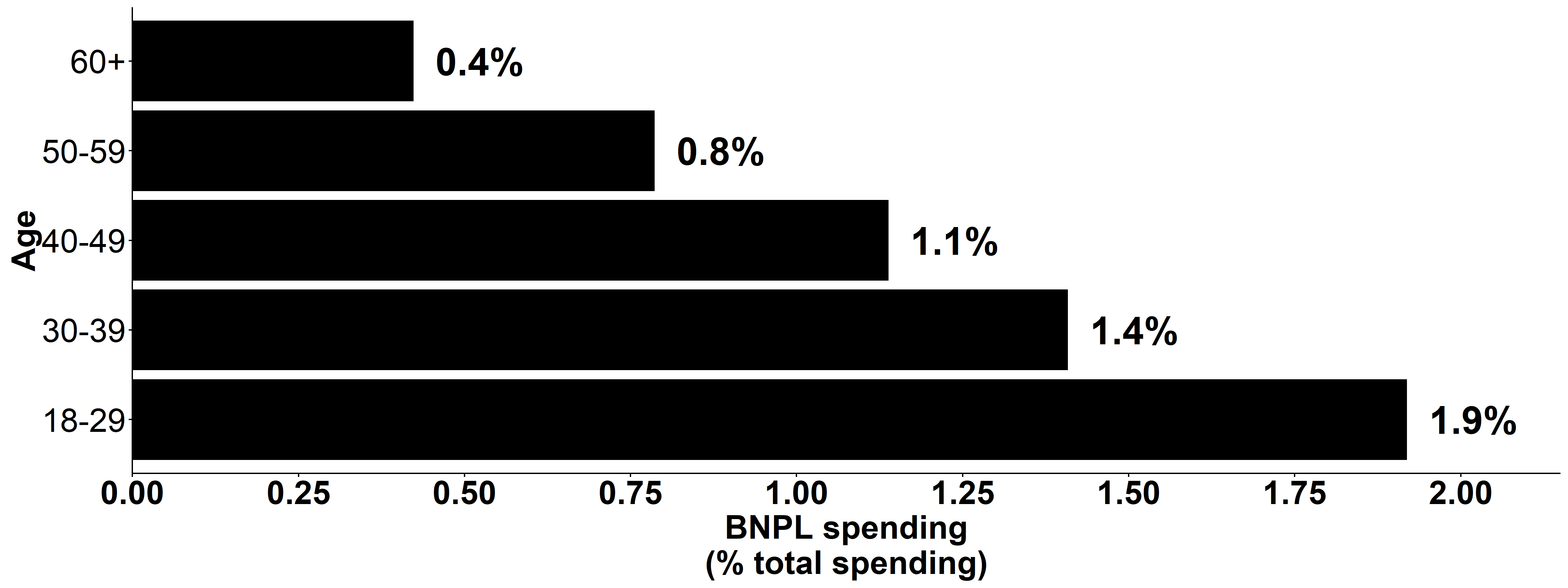}} \\
    \end{tabular}
    \end{center}
    
    \begin{singlespace}
    \noindent {\footnotesize \textit{Notes: UK credit card transactions. BNPL is buy now, pay later. The sample of active credit cards are defined as those with any BNPL or non-BNPL transactions in both January and December 2021. Panel A shows whether any BNPL spending is recorded on an active credit card as a percent of all active credit cards. Panel B shows the value of BNPL spending recorded on active credit cards as a percent of the total value of spending on active credit cards.}}
    \end{singlespace}
\end{figure}

What heterogeneity by consumers' age is there in BNPL usage on credit cards?
Figure \ref{fig:age} finds the use of BNPL on credit cards is more prevalent among younger consumers.
84\% of overall transactions by BNPL firms on credit cards is by consumers aged 18 to 49.
While we do not observe credit card repayments or interest in our data, we know more generally that younger credit cardholders in the UK are least likely to repay their credit card balances in full and so are more likely to incur interest costs.\footnote{Financial Conduct Authority's (FCA) nationally-representative UK Financial Lives survey reports the proportion of UK consumers who revolve credit card debt (conditional on holding any credit card) by age: 30\% of 18-24, 46\% of 25-34, 46\% of 35-44 to 37\% of 45-55, 22\% of 55-64, 10\% of 65+. This survey does not contain any information on BNPL usage or the number of credit cards held preventing a like-for-like comparison. See \cite{fulford2015consumer} for estimates of revolving behavior across the life cycle for US consumers.}

Regulators may be interested in evaluating BNPL's role as gateway debt products for young consumers who are inexperienced users of financial products.
BNPL may potentially be harmfully leading young consumers into a debt spiral of taking on increasingly expensive forms of debt.
Or conversely, BNPL may be beneficially enabling young consumers to learn to prudently use low cost credit, improve their credit access, and avoid relying on higher cost products.

\subsection{BNPL Usage by Area Deprivation}

We seek to understand the vulnerability of consumers using BNPL. 
Regulators' risk assessments consider the vulnerability of consumers using financial products  (e.g. giving higher welfare weights to more deprived consumers).\footnote{\href{https://www.fca.org.uk/publication/consultation/cp21-13.pdf}{https://www.fca.org.uk/publication/consultation/cp21-13.pdf} \& \href{https://www.fca.org.uk/publication/finalised-guidance/fg21-1.pdf}{https://www.fca.org.uk/publication/finalised-guidance/fg21-1.pdf}}
From a consumer welfare standpoint, a regulator may be more worried about potential consumer detriment from a (ultimately) high-interest product sold to lower income, less financially-capable consumers than a zero-interest product sold to consumers wealthy consumers. 
While we cannot observe income and other consumer characteristics at the individual level, we can draw upon postcode identifiers of the cardholder's address to match in measures at the local level.

To evaluate this aspect we aggregate data to the Local Authority District -- areas of local government -- and merge in the 2019 English Indices of Multiple Deprivation (IMD).
IMD is the official government measure providing a relative ranking across districts of deprivation.
IMD is constructed from 39 indicators to ensure it captures a broad range of resources.\footnote{\href{https://assets.publishing.service.gov.uk/government/uploads/system/uploads/attachment\_data/file/833951/IoD2019\_Technical\_Report.pdf}{https://assets.publishing.service.gov.uk/government/uploads/system/uploads/attachment\_data/file/833951/IoD2019\_Technical\_Report.pdf}}

\begin{figure}[!htb]
  \caption{BNPL usage on active credit cards by local area deprivation, 2021}
    \label{fig:lad}
    \begin{center}
    \begin{tabular}{cc} 
    \multicolumn{2}{c}{\textbf{I. BNPL usage on credit cards active in December 2021}} \\
    \textbf{A. BNPL usage (\% credit cards)} & \textbf{B. BNPL usage (\% credit card spending)} \\ 
    {\includegraphics[width=2.75in]{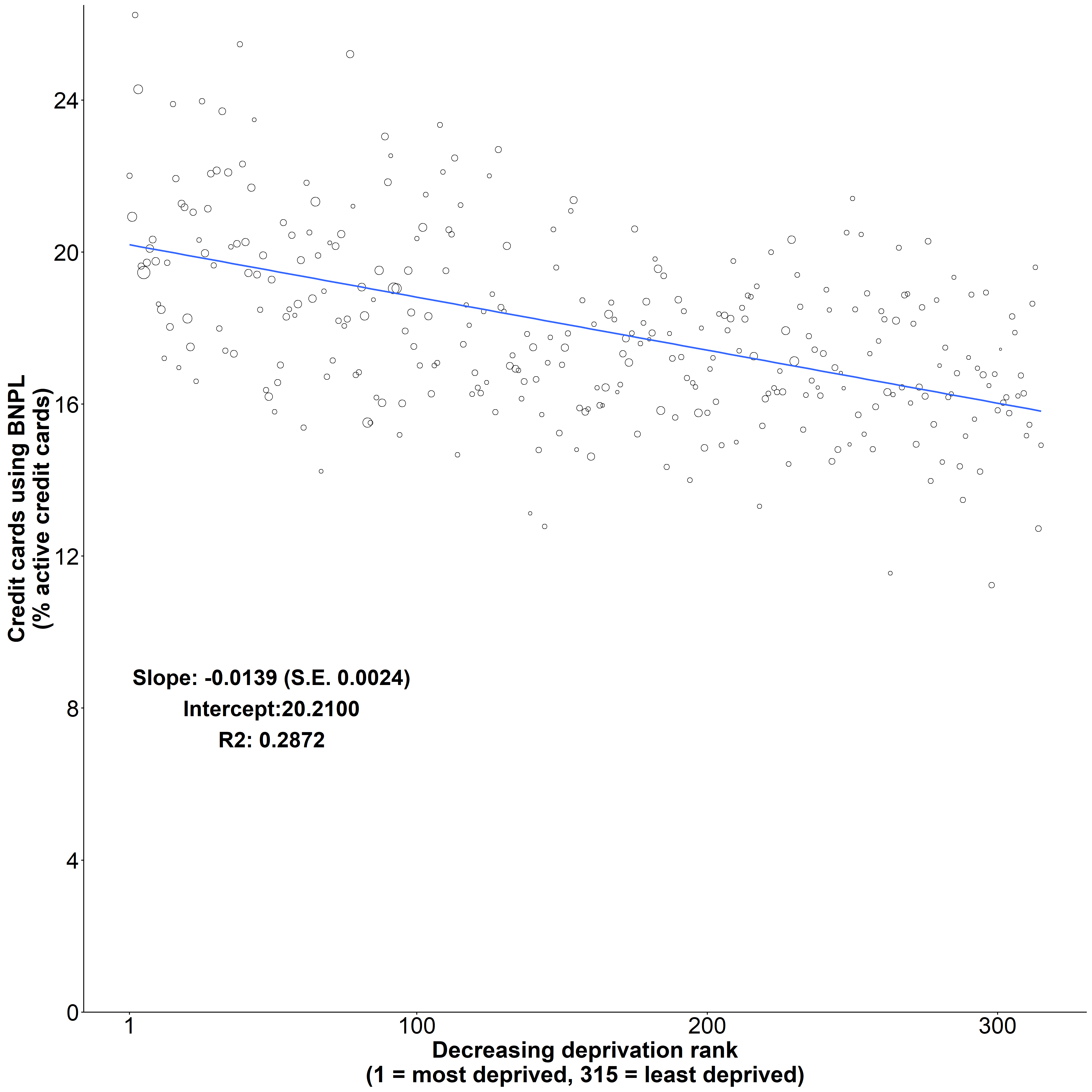}} & {\includegraphics[width=2.75in]{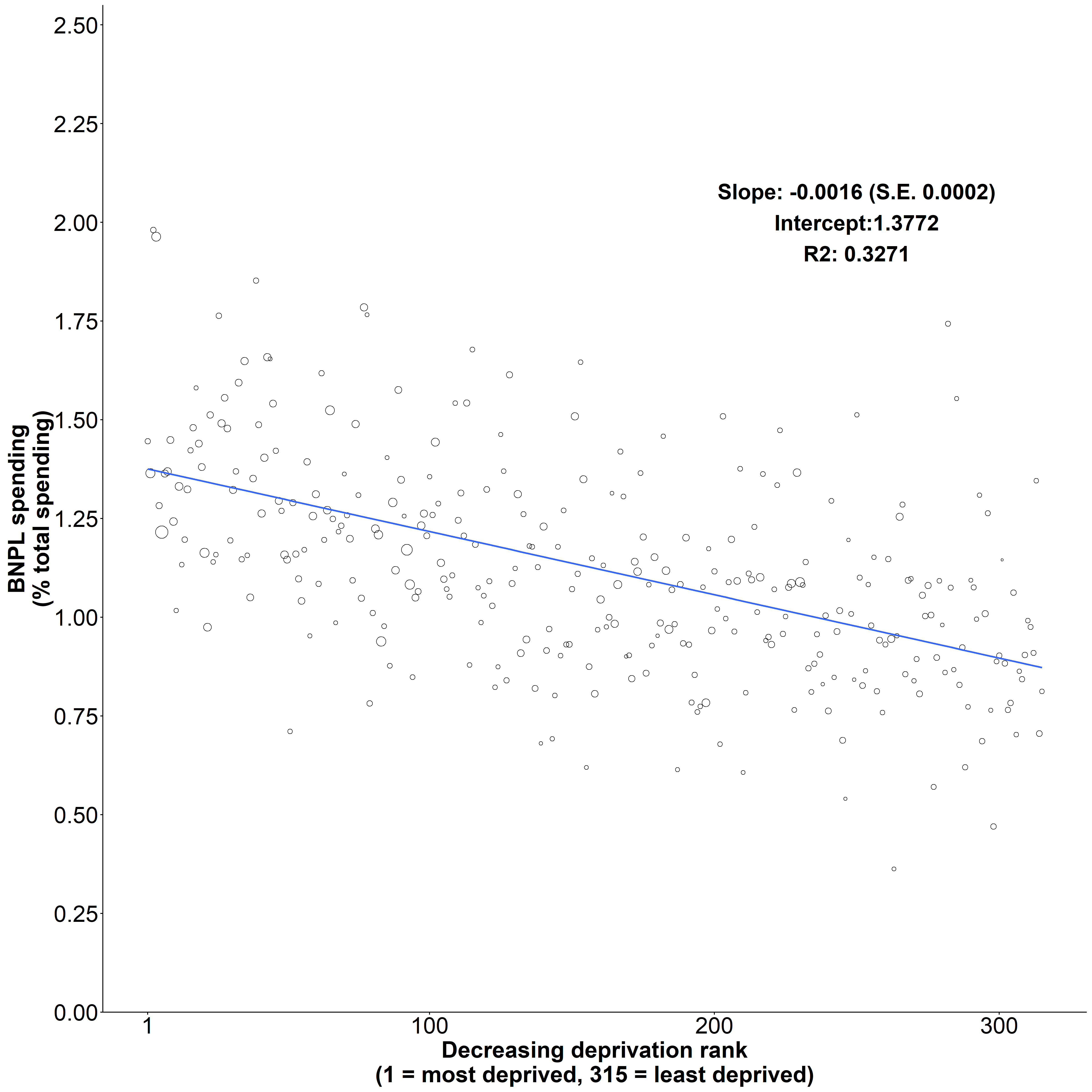}} \\ \\ 
    \multicolumn{2}{c}{\textbf{II. BNPL usage on credit cards active in both January and December 2021}} \\ 
    \textbf{C. BNPL usage (\% credit cards)} & \textbf{D. BNPL usage (\% credit card spending)} \\ 
    {\includegraphics[width=2.75in]{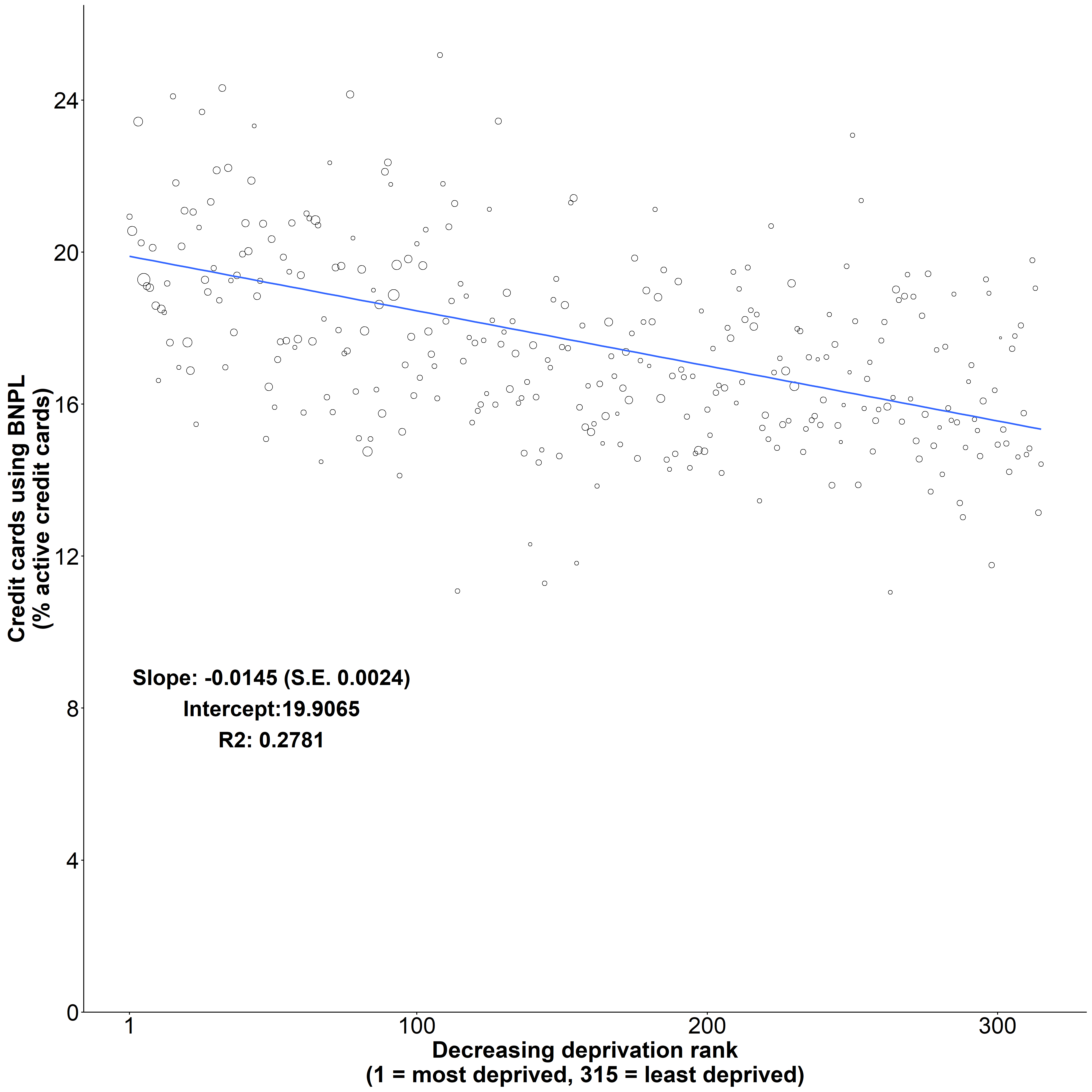}} & {\includegraphics[width=2.75in]{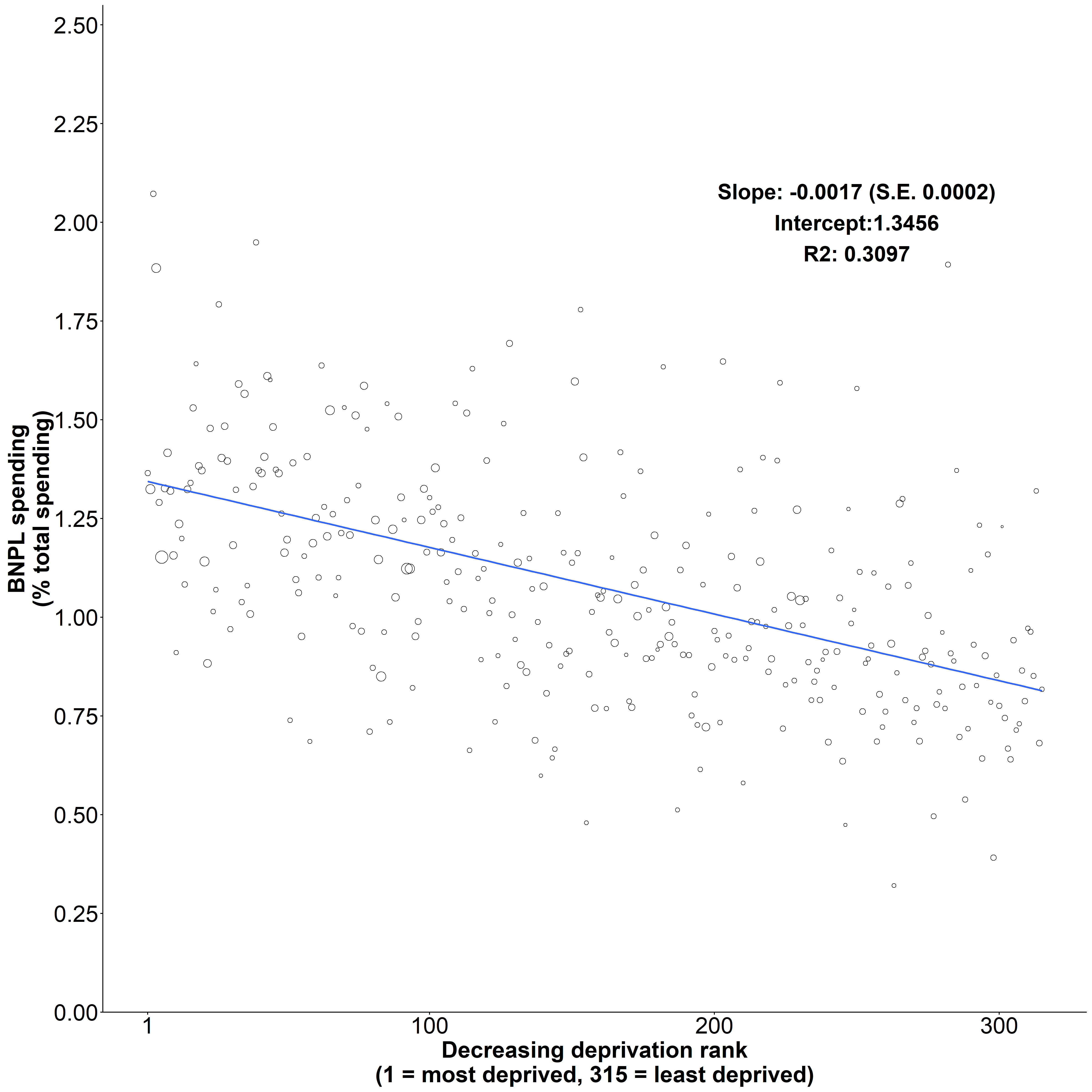}} \\
    \end{tabular}
    \end{center}
    
    \begin{singlespace}
    \noindent {\footnotesize \textit{Notes: Credit card transactions in England, Ministry of Housing, Communities \& Local Government (MHCLG), Office for National Statistics (ONS) data. BNPL is buy now, pay later. Data aggregated to Local Authority District (LAD) level based on cardholder postcode sector. Cardholders in England across 315 LADs since there is no official standardized UK-wide index of multiple deprivation. Due to small populations, City of London is merged with Westminster and Isles of Scilly merged with Cornwall. Size of dot is share of ONS England population estimates and the linear regression is weighted by these shares with standard errors clustered by UK region. Active credit cards are those with any BNPL or non-BNPL spending. Panels A and C are percent of the number of active credit cards in a LAD which have any BNPL spending. Panels B and D are percent of the value of credit card spending in a LAD on BNPL. Deprivation ranks by English Indices of Multiple Deprivation (2019) - more details: \href{www.gov.uk/government/statistics/english-indices-of-deprivation-2019}{www.gov.uk/government/statistics/english-indices-of-deprivation-2019}}}
    \end{singlespace}

\end{figure}

Panels in Figure \ref{fig:lad} rank each authority by its IMD on the x-axes where a ranking of 1 is most deprived and 315 the least.
Panels A and B use the sample of cards active in December 2021 and panels C and D those active in both January and December 2021 (results are consistent).
We measure BNPL usage on credit cards the y-axes: Panels A and C use the percent of credit cards with \textit{any} transaction by a BNPL firm in the last 12 months, Panels B and D use the share of the total value of spending on credit cards that by a BNPL firm in the last 12 months.
The size of each dot is each district's share of population.

We plot a linear regression (Equation 2) to describe the unconditional (non-causal) relationship between IMD and BNPL usage.
There is one observation ($d$) per district in England (315 in total) where $Y_d$ are our measures of BNPL usage (0-100) and $IMD_d$ is the ranking of IMD (1-315 where a higher value is \textit{less} deprived).
We weight each observation by its district's ONS population and cluster standard errors by regions of the UK to allow for spatial correlation across districts.

\begin{equation}
    Y_d = \alpha + \beta \; IMD_d + \varepsilon_d
\end{equation}

By both measures of BNPL usage we find the most deprived areas have statistical significantly (p-values $<$ 0.001), higher charging of BNPL to credit cards in more deprived areas.
The $\beta$ coefficients can be interpreted as follows: moving from the least to most deprived local authority is associated, on average, with \textit{higher} BNPL use: 28 to 30\% (4.4 to 4.6 pp) as measured by fraction of cards, and 58 to 66\% (0.5 pp) measured by the fraction of spending.\footnote{Results robust to using income instead of IMD: BNPL use is 30\% higher as measured by fraction of cards, and 68\% higher as measured by the fraction of spending moving from the highest to lowest income local authority.}
Given this relationship it is unclear how much BNPL lenders restrict credit supply - something visible in regulated credit (e.g. \citealp{agarwal2018banks}).
While we do not observe whether these credit cards are repaid in full or accumulate fees or interest (or at what rate), we know more generally that credit cards in the most deprived areas of England are those least likely to repay in full, be eligible for 0\% credit card deals, and instead have higher interest rates.
\footnote{Financial Conduct Authority's nationally-representative Financial Lives survey reports the proportion of UK consumers who revolve credit card debt (conditional on holding any credit card) is 52\% for the most deprived decile of IMD compared to 20\% in the least deprived decile of IMD. This survey does not contain any information on BNPL usage or the number of credit cards held preventing a like-for-like comparison.}

While this relationship indicates BNPL's potential risks for regulators to consider, its welfare effects are unclear.
BNPL may be welfare improving enabling consumption smoothing for liquidity-constrained consumers in poorer areas with high MPCs to increase their consumption or use BNPL to save money instead of borrowing on higher cost credit (e.g. overdrafts, payday loans).
However, BNPL may be welfare decreasing if BNPL leads to increased borrowing on higher cost credit or consumers in poorer areas are less financially sophisticated or making na\"ive mistakes (e.g. overoptimism and/or present focused in their ability to repay BNPL debt). 
The effects of such na\"ivete has previously been evaluated in credit cards \citep[e.g.][]{meier2010present} and payday loans \citep[e.g.][]{allcott2021high}.
By increasing indebtedness, BNPL may cause broader distress (e.g. adverse mental health, missing other bills) that may outweigh consumption benefits.

\section{Conclusions}

We document that a large minority of UK consumers charge BNPL to credit cards, especially younger consumers and those living in in more deprived areas. 
This raises doubts on these consumers' ability to pay for BNPL and prompts a regulatory question as to whether consumers should be allowed to refinance their unsecured debt.
Further research is required to measure how binding liquidity constraints are for consumers using BNPL, the effects of BNPL on consumers, whether BNPL is substituting for other payment mechanisms or forms of lending, and BNPL's competitive effects on these established products. 
Given the current interest by policymakers in the BNPL market, further research could fruitfully evaluate the effects of potential BNPL regulations.

\clearpage

\section{Figures}

\clearpage

\clearpage

\clearpage

\appendix

\begin{center}
\Large
\noindent \textbf{Online Appendix for ``Buy Now, Pay Later (BNPL)...On Your Credit Card''} \\ 
\end{center}

\section{Other Sources of Evidence on BNPL}

We conducted a literature review in December 2021 before publicly releasing the first version of our working paper on ArXiv in January 2022.
At the time of writing this paper, searches for `Buy Now, Pay Later' and `BNPL' on ArXiv returned no results; on SSRN returned a single law working paper on the Singapore BNPL market \citep{sng2021buy}; there were no relevant results on NBER Working Papers beyond \cite{berg2021fintech}'s broader review of FinTech lending, and on google scholar showed no economics or finance papers in either published or working form (the only work was from other academic fields (\cite{xing2019australian}, \cite{fook2020click}, \cite{gerrans2021fintech}, \cite{johnson2021analyzing}).
In terms of other sources of evidence there are a few consumer surveys, blog discussions, and industry reports.\footnote{Citizens Advice UK surveys: \href{https://www.citizensadvice.org.uk/about-us/our-work/policy/policy-research-topics/debt-and-money-policy-research/buy-nowpain-later/}{https://www.citizensadvice.org.uk/about-us/our-work/policy/policy-research-topics/debt-and-money-policy-research/buy-nowpain-later/} and  \href{https://www.citizensadvice.org.uk/about-us/our-work/policy/policy-research-topics/debt-and-money-policy-research/buy-now-pay-later-what-happens-if-you-cant-pay-later/}{https://www.citizensadvice.org.uk/about-us/our-work/policy/policy-research-topics/debt-and-money-policy-research/buy-now-pay-later-what-happens-if-you-cant-pay-later/}, Which? UK surveys:
\href{https://www.which.co.uk/policy/money/7601/buynowpaylater}{https://www.which.co.uk/policy/money/7601/buynowpaylater} and 
\href{https://www.which.co.uk/policy/money/8573/buynowpaylater2}{https://www.which.co.uk/policy/money/8573/buynowpaylater2}, Money.co.uk UK survey: \href{https://www.money.co.uk/guides/generation-debt-trap}{https://www.money.co.uk/guides/generation-debt-trap} and \cite{battermann21} qualitative study.
Credit Karma US survey: \href{https://www.creditkarma.com/about/commentary/buy-now-pay-later-surges-throughout-pandemic-consumers-credit-takes-a-hit}{https://www.creditkarma.com/about/commentary/buy-now-pay-later-surges-throughout-pandemic-consumers-credit-takes-a-hit} \& Federal Reserve surveys:
\href{https://www.federalreserve.gov/consumerscommunities/shed.htm}{https://www.federalreserve.gov/consumerscommunities/shed.htm} \& \href{https://www.philadelphiafed.org/-/media/frbp/assets/consumer-finance/discussion-papers/dp22-02.pdf}{https://www.philadelphiafed.org/-/media/frbp/assets/consumer-finance/discussion-papers/dp22-02.pdf}
For examples of US blogs on BNPL see \cite{alcazar21a,alcazar21b,Akeredolu21,lott21} and Equifax's report \href{https://insight.equifax.com/what-to-know-about-buy-now-pay-later/}{https://insight.equifax.com/what-to-know-about-buy-now-pay-later/}
For examples of industry analysis see Accenture US BNPL market report commissioned by Afterpay:
\href{https://afterpay-corporate.yourcreative.com.au/wp-content/uploads/2021/10/Economic-Impact-of-BNPL-in-the-US-vF.pdf}{https://afterpay-corporate.yourcreative.com.au/wp-content/uploads/2021/10/Economic-Impact-of-BNPL-in-the-US-vF.pdf} and
Bain UK BNPL market report:
\href{https://www.bain.com/globalassets/noindex/2021/bain\_report\_buy\_now\_pay\_later-in-the-uk.pdf}{https://www.bain.com/globalassets/noindex/2021/bain\_report\_buy\_now\_pay\_later-in-the-uk.pdf}.}

\section{Institutional Details on BNPL}

Economic theory provides mixed perspectives on the consumer benefits of deferred payments such as BNPL.
Through the lens of the canonical life-cycle model of consumption smoothing, opportunities to smooth consumption at zero interest cost are weakly welfare improving -- especially in a high inflation environment \citep{ando1963life}.
Although deferring payments allows consumption smoothing benefits, such behavior can also present the possibility for welfare losses -- for example by financially unsophisticated or na\"ive present focused consumers mistakenly overconsuming \citep[e.g.][]{allcott2021high}.

BNPL is a FinTech credit product offered at the point of sale providing consumers the option to pay for their purchase at a later date in one or more interest-free instalments \citep{hmt21,cfpb21}.
BNPL is mainly used online with lenders typically a third party separate to the retailer. 

Consumers are (typically) charged no interest or fees \textit{unless} they miss BNPL payments.
They may experience costs if their BNPL payment has knock-on adverse effects on their other finances (e.g. triggering overdraft fees, accumulating credit card interest, missing payments on other household bills).
Such increased broader financial distress from taking on credit was previously found in the UK payday lending market \citep{gathergood2019payday}.

Repayment structures vary across and within UK BNPL lenders: for example, Klarna provides an option to repay in the next thirty days as well as an option to repay in three instalments thirty days apart whereas Clearpay has four payments that are two weeks apart.
UK and UK BNPL product offerings are similar -- often offered by the same global lenders.
The \cite{cfpb21} defined a BNPL as having 4 or fewer instalments but some BNPL products have more (e.g. OpenPay in the UK offers a choice of 3-10 instalments).

Although leading BNPL firms operate an instalment model, FinTech product innovation means the distinction between instalments and revolving credit (e.g. credit cards and retail store cards) is not clear cut.
New BNPL products are emerging (e.g. Instalments by Barclays for Amazon UK, the Affirm Card is a debit card with BNPL) which feature a credit limit and so have some similarities to  historical retail (store) cards.
Some credit card providers offer the option for cardholders to pay particular transactions in BNPL-esque instalments (e.g. American Express's `Pay It Plan It' and Barclaycard's Instalment Plan) as do some bank (current / checking) account providers on their debit cards (e.g. Monzo Flex).

In the BNPL market not all firms charge late fees (e.g. PayPal and Klarna UK do not).
When BNPL firms do charge late fees (e.g. Afterpay/Clearpay and Klarna US) they are smaller than those on other credit products such as credit cards where the UK industry standard fee is \pounds12 plus additional interest costs.
Non-payment of BNPL debt can still have consequences: BNPL lenders may block new purchases by that consumer, pass unpaid debt to debt collectors, and missed payments may get reported in credit files, however, the prevalence and effects of such practices are unknown.
US survey evidence estimates a third of BNPL users have missed payments and those consumers report their credit scores having since declined, however, this does not provide causal evidence of BNPL's effects on credit scores -- indeed Equifax FICO analysis (of the very small, selected subset of BNPL reported in credit files) indicates mixed impacts on credit scores.\footnote{\href{https://www.creditkarma.com/about/commentary/buy-now-pay-later-surges-throughout-pandemic-consumers-credit-takes-a-hit}{https://www.creditkarma.com/about/commentary/buy-now-pay-later-surges-throughout-pandemic-consumers-credit-takes-a-hit} \href{https://www.equifax.com/resource/-/asset/webinar/market-pulse-buy-now-pay-later-credit-score-impact-analysis-webinar/}{https://www.equifax.com/resource/-/asset/webinar/market-pulse-buy-now-pay-later-credit-score-impact-analysis-webinar/}}
The practice of unregulated BNPL lenders charging consumers fees in the US has been challenged leading to a series of lenders agreeing to refund consumers \$1.9 mn in settlements with Californian financial regulators \citep{alcazar21b}.\footnote{\href{https://dfpi.ca.gov/2021/10/07/dfpi-report-shows-changes-in-consumer-lending-decrease-in-pace-program/}{https://dfpi.ca.gov/2021/10/07/dfpi-report-shows-changes-in-consumer-lending-decrease-in-pace-program/}}

Merchant fees appear the primary source of revenue for BNPL lenders.
Retailers offer BNPL in exchange for merchant fees that can be 3-6\% and gaining insights about their customers from BNPL lenders to assist with targeted marketing  \citep{cfpb21}.
It is estimated the gross profit margins of BNPL lenders is thirty basis points.\footnote{\href{https://www.ft.com/content/ddb2e207-2450-4ca8-bad0-871290d80ea7}{https://www.ft.com/content/ddb2e207-2450-4ca8-bad0-871290d80ea7}}
By revealed preference, BNPL is generating more net revenue in additional retail sales for these retailers than the fees they are giving up: by one estimate these increase conversation rates 20-30\% and average transaction size 30-50\%.\footnote{\href{www.cnbc.com/2021/09/25/why-retailers-are-embracing-buy-now-pay-later-financing-services.html}{https://www.cnbc.com/2021/09/25/why-retailers-are-embracing-buy-now-pay-later-financing-services.html}, \href{https://www.klarna.com/us/blog/why-use-buy-now-pay-later-klarna/}{https://www.klarna.com/us/blog/why-use-buy-now-pay-later-klarna/}, \href{https://afterpay-corporate.yourcreative.com.au/wp-content/uploads/2021/10/Economic-Impact-of-BNPL-in-the-US-vF.pdf}{https://afterpay-corporate.yourcreative.com.au/wp-content/uploads/2021/10/Economic-Impact-of-BNPL-in-the-US-vF.pdf}, \href{https://www.bain.com/globalassets/noindex/2021/bain\_report\_buy\_now\_pay\_later-in-the-uk.pdf}{https://www.bain.com/globalassets/noindex/2021/bain\_report\_buy\_now\_pay\_later-in-the-uk.pdf}}
One potential way in which BNPL can increase sales is enhancing the shopping experience by enabling liquidity-constrained consumers to purchase excess amounts of clothing online -- more than their cash balances allow -- to try on and then return the ones they do not want: with BNPL they will typically not be charged for these returns, unlike with traditional instant payments.
Another way BNPL may be profitable is by driving increased, potentially impulsive, merchant sales consumers may regret: one anecdotal example are advertisements encouraging people to purchase cake and pizzas (that only cost a few pounds) on BNPL.\footnote{\href{https://www.ft.com/content/c4da9b2f-5187-4956-931d-0554d4268d4e}{https://www.ft.com/content/c4da9b2f-5187-4956-931d-0554d4268d4e}}

In the UK and US, most BNPL is unregulated \citep{cfpb21,woolard21,hmt21}.\footnote{Most BNPL is not subject to the UK's Consumer Credit Act (CCA) or Financial Services \& Markets Act (FSMA).}
In the context of the UK market this means, unlike regulated credit, BNPL lenders are not required to provide pre-contractual information disclosures, are exempt from advertising rules on financial promotions, do not have a regulatory requirement to conduct an assessment of whether the applicant can afford such credit, and consumers are ineligible to claim redress or make complaints appealing to the financial services ombudsman.

Rapid growth is present in the US market based on limited data available: the California financial regulator reported 530.2\% and 96.8\% increases between 2019 and 2020 in the number and value of finance loans respectively: 91\% of these loans are BNPL.\footnote{\href{https://dfpi.ca.gov/2021/10/07/dfpi-report-shows-changes-in-consumer-lending-decrease-in-pace-program/}{https://dfpi.ca.gov/2021/10/07/dfpi-report-shows-changes-in-consumer-lending-decrease-in-pace-program/}}

Klarna is the BNPL lender with the highest value of UK transactions on credit cards throughout 2019 to 2021 in our data (in line with anecdotal evidence on it being the leading UK BNPL lender).
Clearpay (known as Afterpay in the US) is second, with approximately a quarter of the BNPL transactions value -- a third of Klarna's -- as of December 2021. 
The remaining other BNPL lenders (our other category includes openpay, dividebuy, laybuy and payl8r) appear small, however, as a reminder we caveat that our analysis does not include PayPal's BNPL arm.
We infer that other BNPL lenders not observed in our data are either too small to be present and/or do not have transactions on credit cards.

Beyond BNPL, our research is a proof of concept for how researchers and consumer financial protection regulators can use real-time household financial transaction data to monitor consumer use and risks of unregulated consumer financial products such as BNPL over time.

\clearpage

\section{Supplementary Figures}

\noindent \textbf{Figure A1. Active credit cards during 2021 by region in real-time data compared to shares of adult population} \\ 
\begin{center}
{\includegraphics[width=6in]{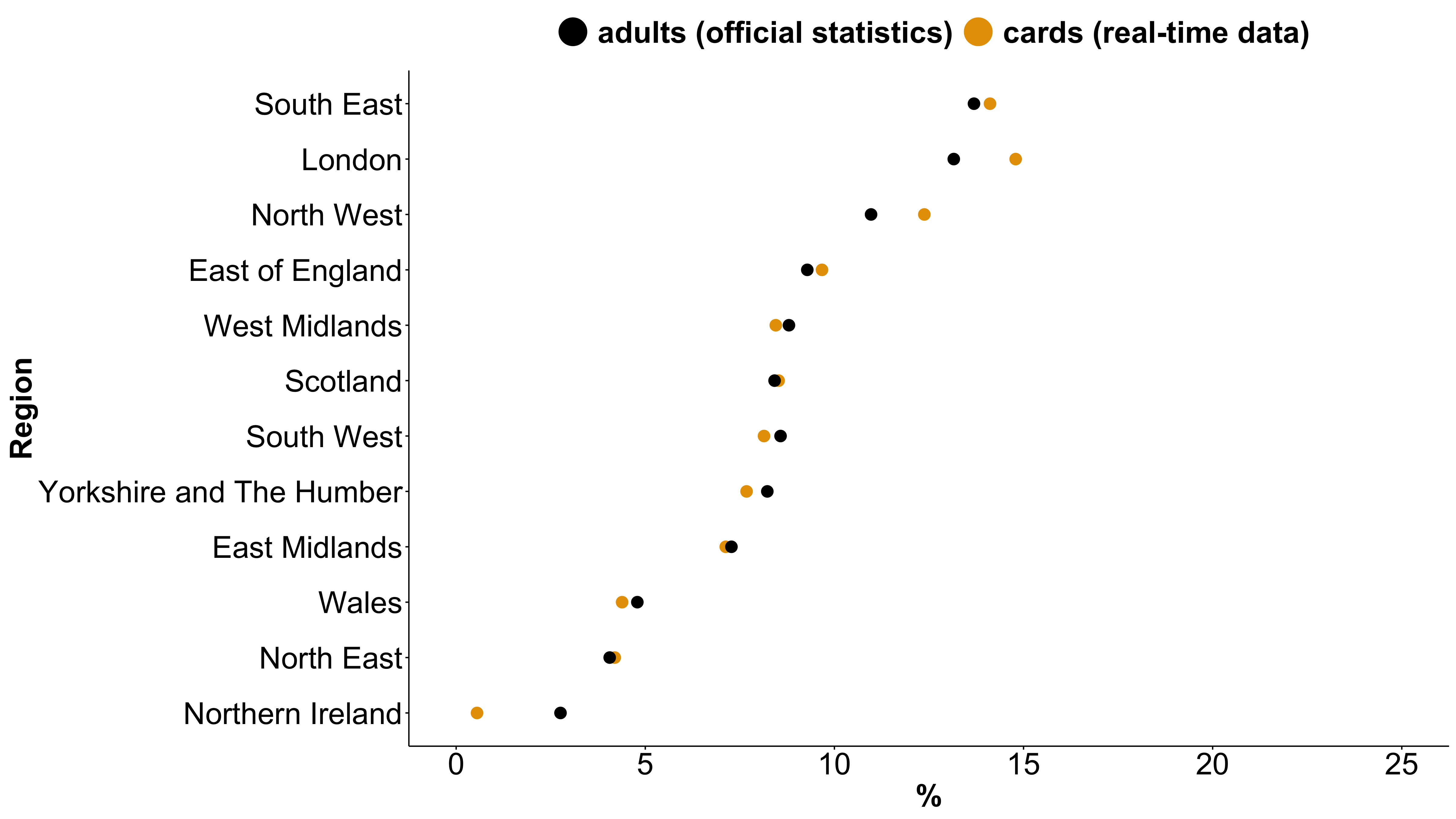}} \\
\end{center}

\begin{singlespace}
\noindent {\footnotesize \textit{Notes: Cards (real-time data) is percent of active cards by region in 2021 observed in UK credit card transactions data. Adults (official statistics) is percent of adults by region from Office for National Statistics (ONS) mid-year population estimates for 2019 - 2020.}}
\end{singlespace}

\clearpage

\noindent \textbf{Figure A2. Active credit cards with any BNPL transactions during 2021} \\ 
\begin{center}
{\includegraphics[width=6in]{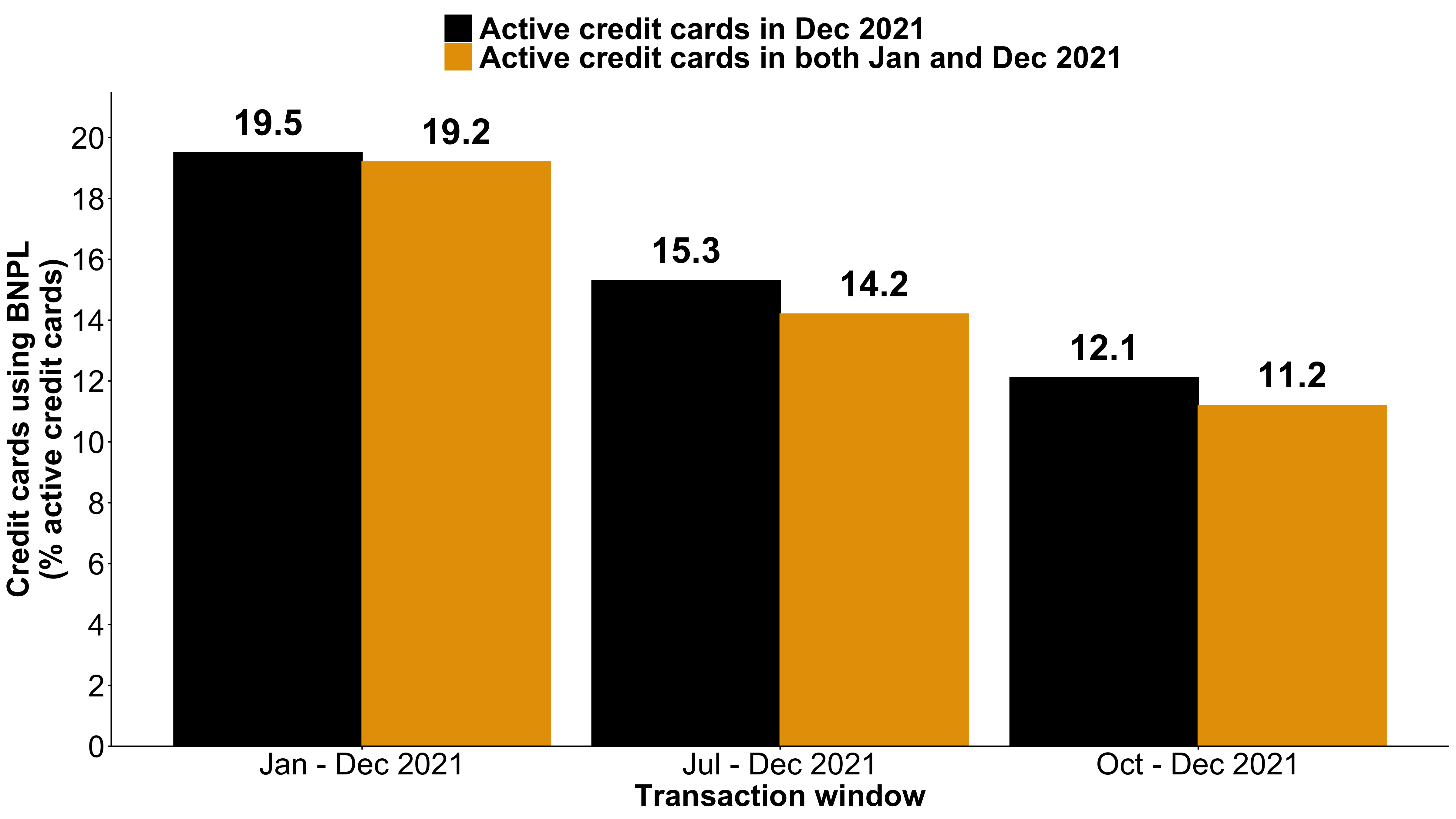}} \\
\end{center}

\begin{singlespace}
\noindent {\footnotesize \textit{Notes: UK credit card transactions data. BNPL is buy now, pay later. The sample of active credit cards are defined as those with any BNPL or non-BNPL transactions in December 2021 (black) or in both January and December 2021 (orange).}}
\end{singlespace}

\noindent \textbf{Figure A3. BNPL lender shares of BNPL transactions on active credit cards, 2019 - 2022} \\ 

\begin{center}
\begin{tabular}{c} 
{\includegraphics[width=5.25in]{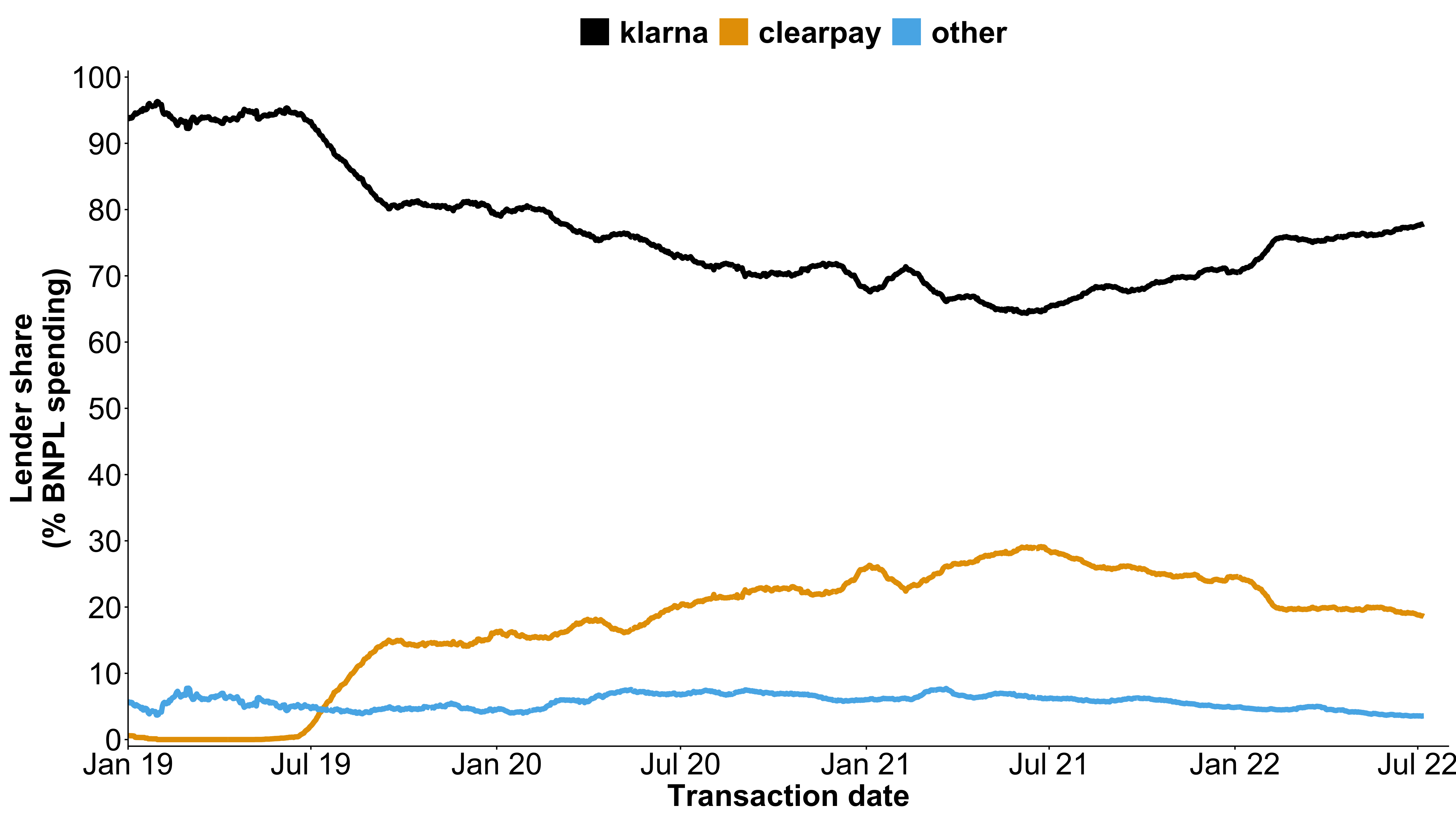}} \\
\end{tabular}
\end{center}

\begin{singlespace}
\noindent {\footnotesize \textit{Notes: UK credit card transactions (repeated cross-section) data. BNPL is buy now, pay later. 28 day moving averages for each BNPL lender's share of the total value of BNPL spending on credit cards.}}
\end{singlespace}

\clearpage

\noindent \textbf{Figure A4. BNPL spending as \% of all spending, 2020 - 2022 (28 day moving average)} \\ 

\begin{center}
\begin{tabular}{c} 
{\includegraphics[width=4in]{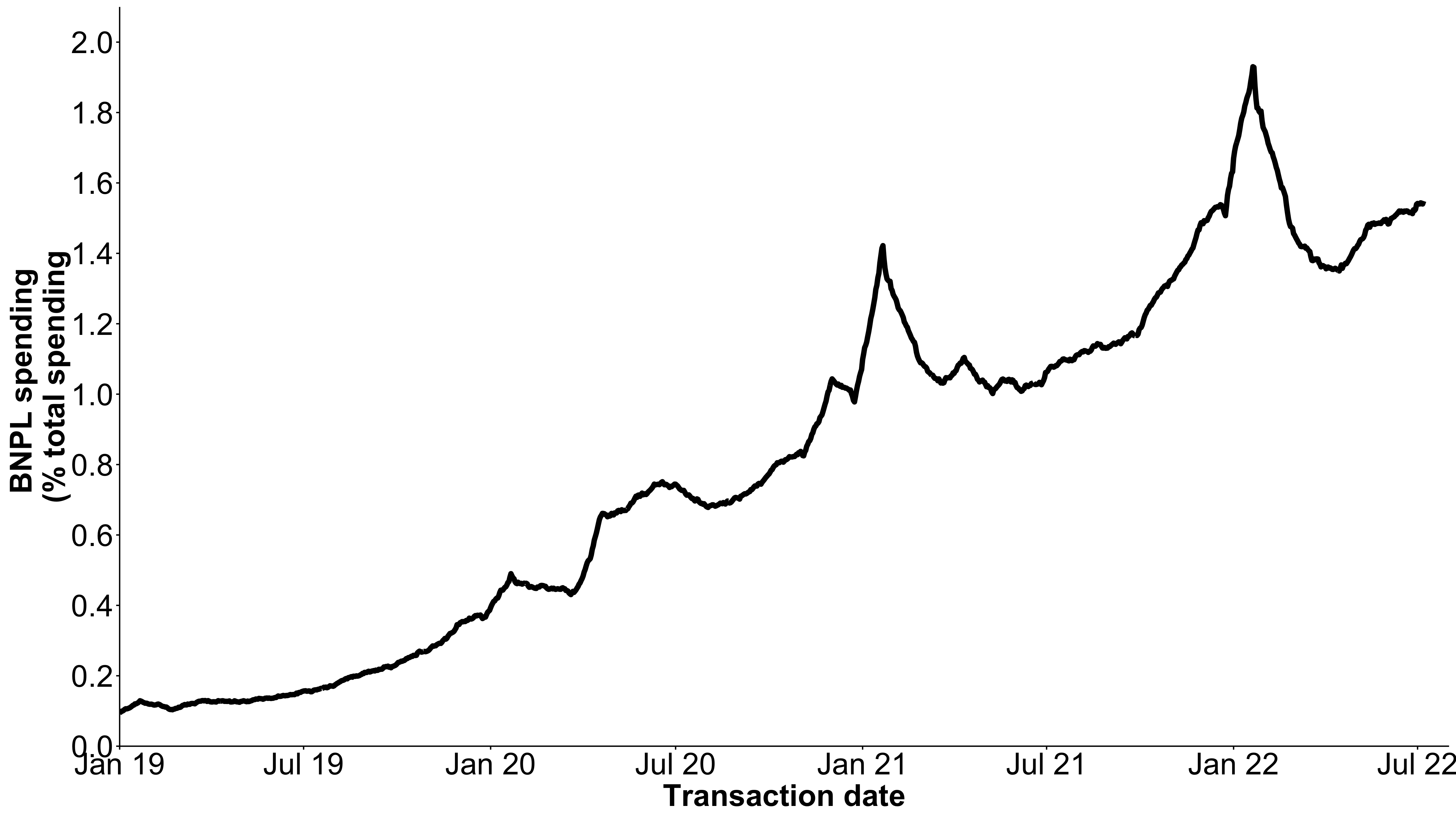}} \\
\end{tabular}
\end{center}

\begin{singlespace}
\noindent {\footnotesize \textit{Notes: UK credit card transactions data (repeated cross section) data. BNPL is buy now, pay later. 28 day moving averages.}} \\
\end{singlespace}

\vspace{-0.1in}

\noindent \textbf{Figure A5. BNPL spending as \% of all spending by age, 2020 - 2021} \\ 

\begin{center}
\begin{tabular}{c} 
{\includegraphics[width=5in]{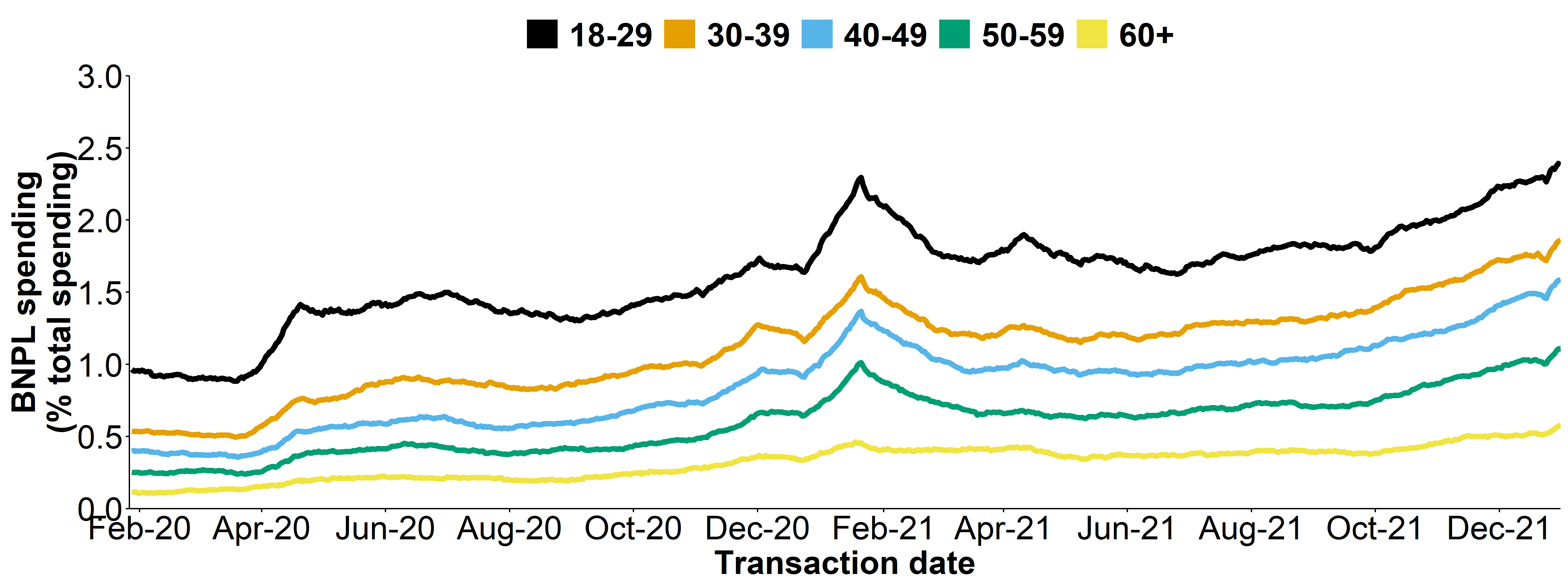}} \\ 
\end{tabular}
\end{center}

\begin{singlespace}
\noindent {\footnotesize \textit{Notes: UK credit card transactions (repeated cross-section) and Office for National Statistics (ONS) data. BNPL is buy now, pay later.}} \\ 
\end{singlespace}

\vspace{-0.1in}

\noindent \textbf{Figure A6. Share of total BNPL spending in 2021 by age} \\ 

\begin{center}
\begin{tabular}{c} 
{\includegraphics[width=4in]{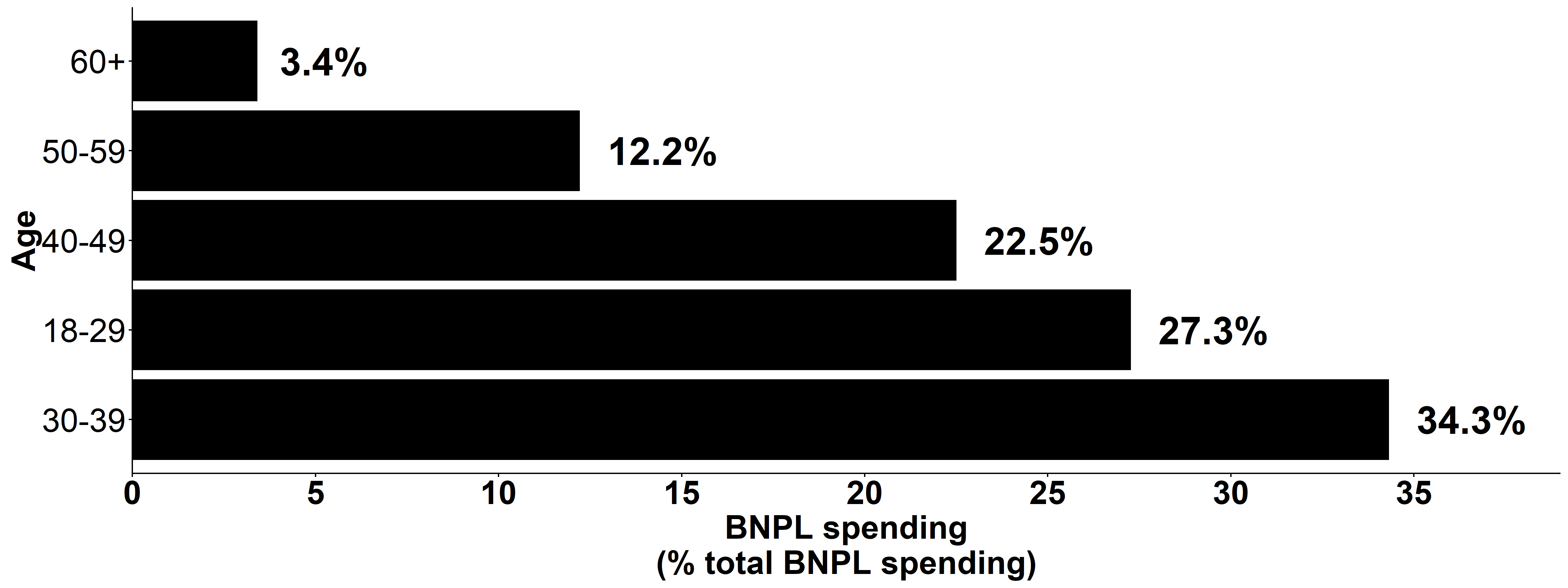}}
\end{tabular}
\end{center}

\begin{singlespace}
\noindent {\footnotesize \textit{Notes: UK credit card transactions (repeated cross-section) data. BNPL is buy now, pay later. Numbers do not sum to 100\% due to rounding.}}
\end{singlespace}

\clearpage

\noindent \textbf{Figure A7. Active credit cards (in December 2021) with any BNPL transactions during 2021, by age (A), region (B), and area (C)} 

\begin{center}
\begin{tabular}{c} 
\textbf{A. Age} \\ 
{\includegraphics[width=3.5in]{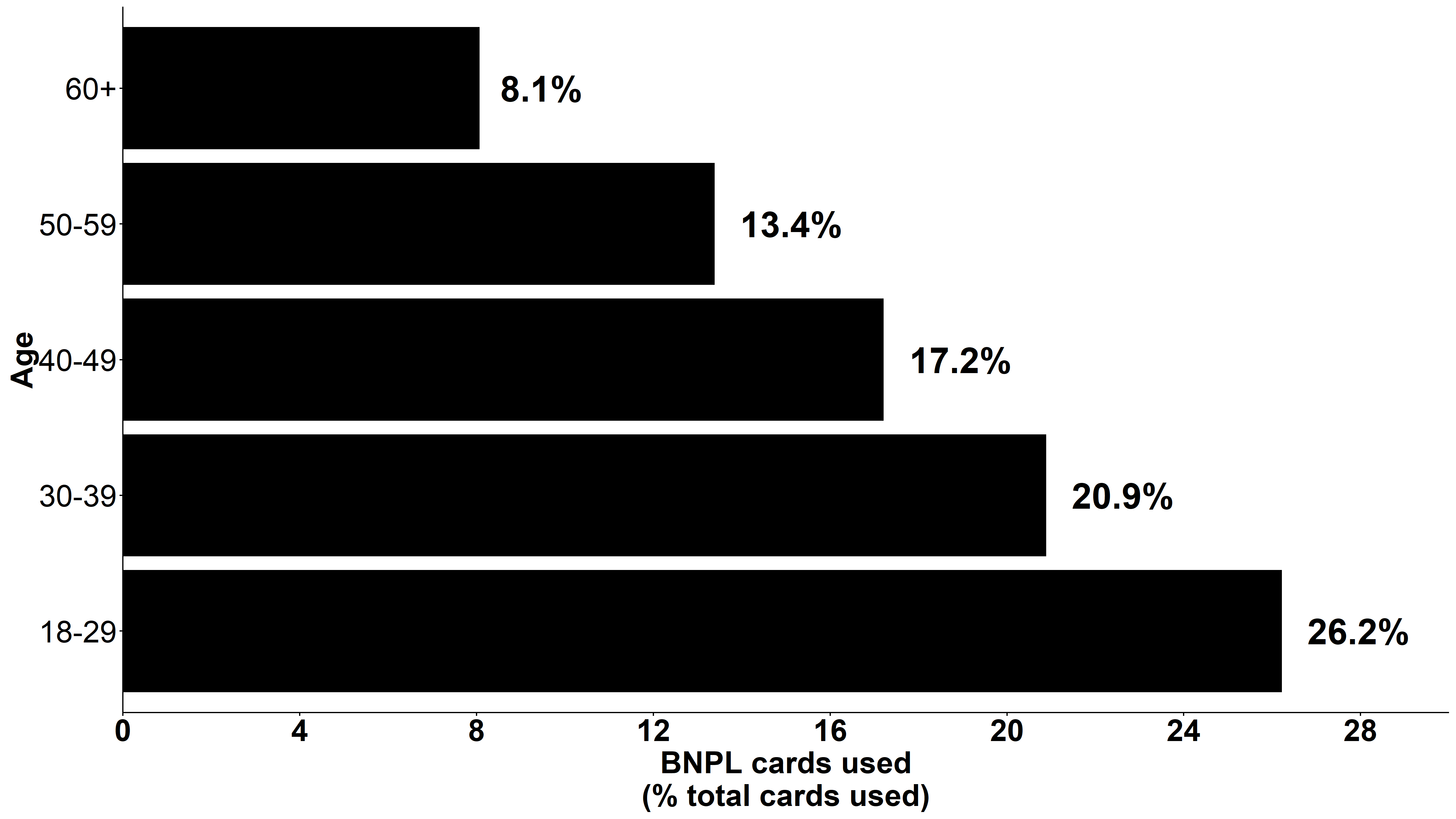}} \\ 
\textbf{B. Region} \\ 
{\includegraphics[width=3.5in]{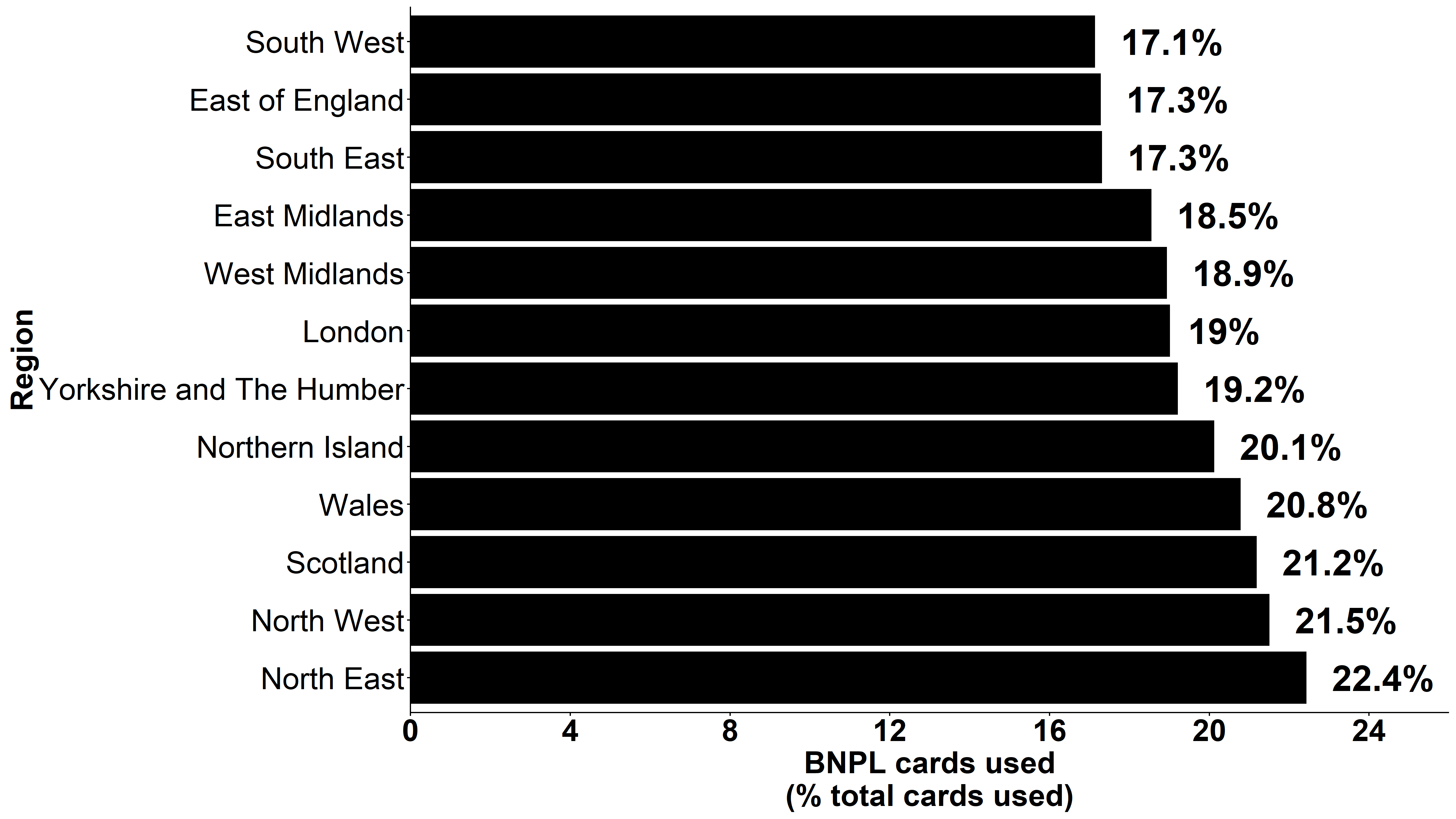}} \\ 
\textbf{C. Urban-rural area classification} \\ 
{\includegraphics[width=3.5in]{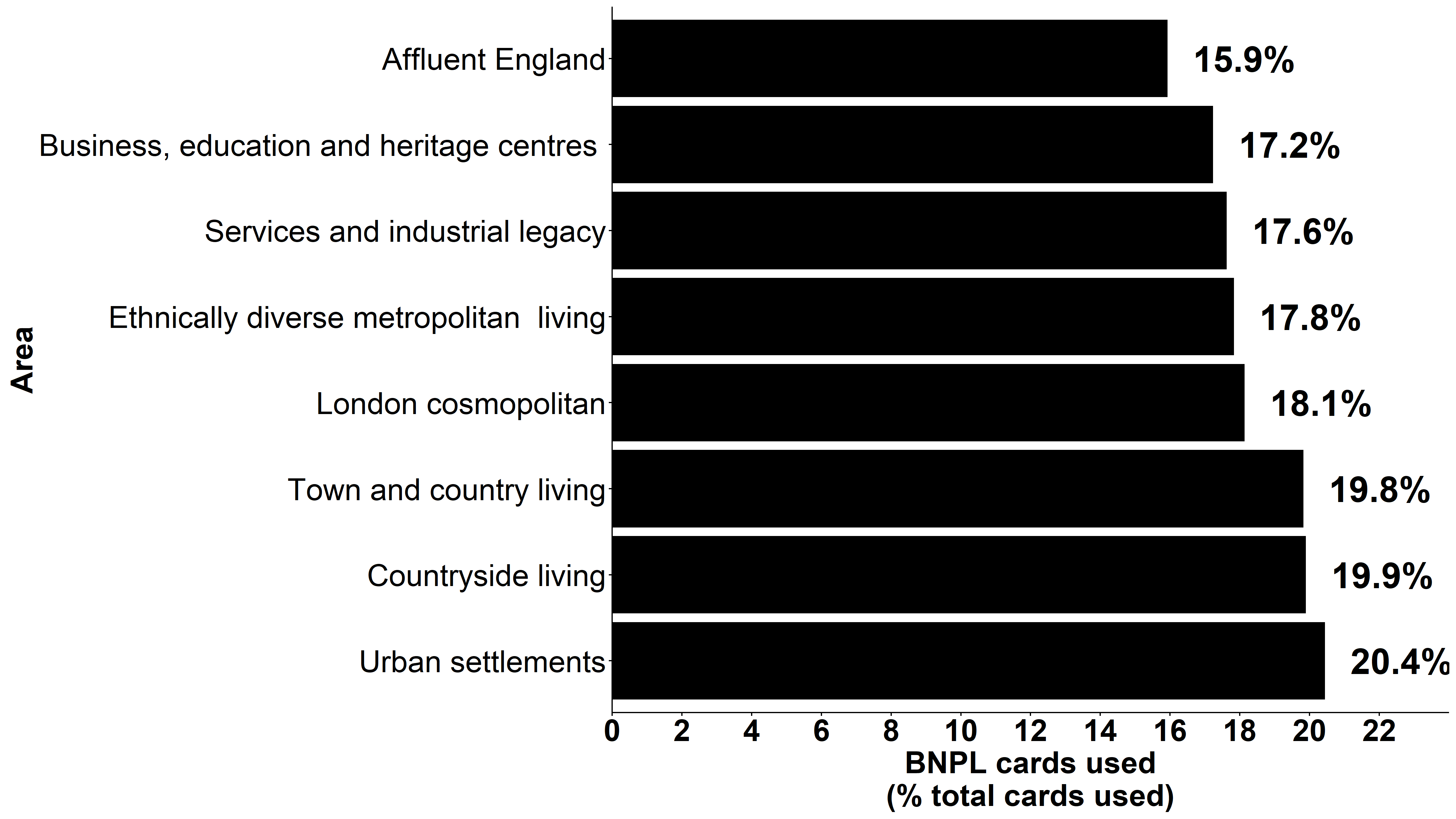}} \\ 
\end{tabular}
\end{center}

\begin{singlespace}
\noindent {\footnotesize \textit{Notes: UK credit card transactions and Office for National Statistics (ONS) data. BNPL is buy now, pay later. Actively-used credit cards defined as any BNPL or non-BNPL transactions in December 2021. Panels B-C allocate cards based on cardholder postcode sector to use ONS NUTS1 regions (Panel B) and ONS supergroup (2011) area classifications  (Panel C). Maps in Figure A8.}}
\end{singlespace}

\clearpage

\noindent \textbf{Figure A8. Active credit cards (in both January and December 2021) with any BNPL transactions during 2021, by region (A), and area (B)} 

\begin{center}
\begin{tabular}{c} 
\textbf{A. Region} \\ 
{\includegraphics[width=3.5in]{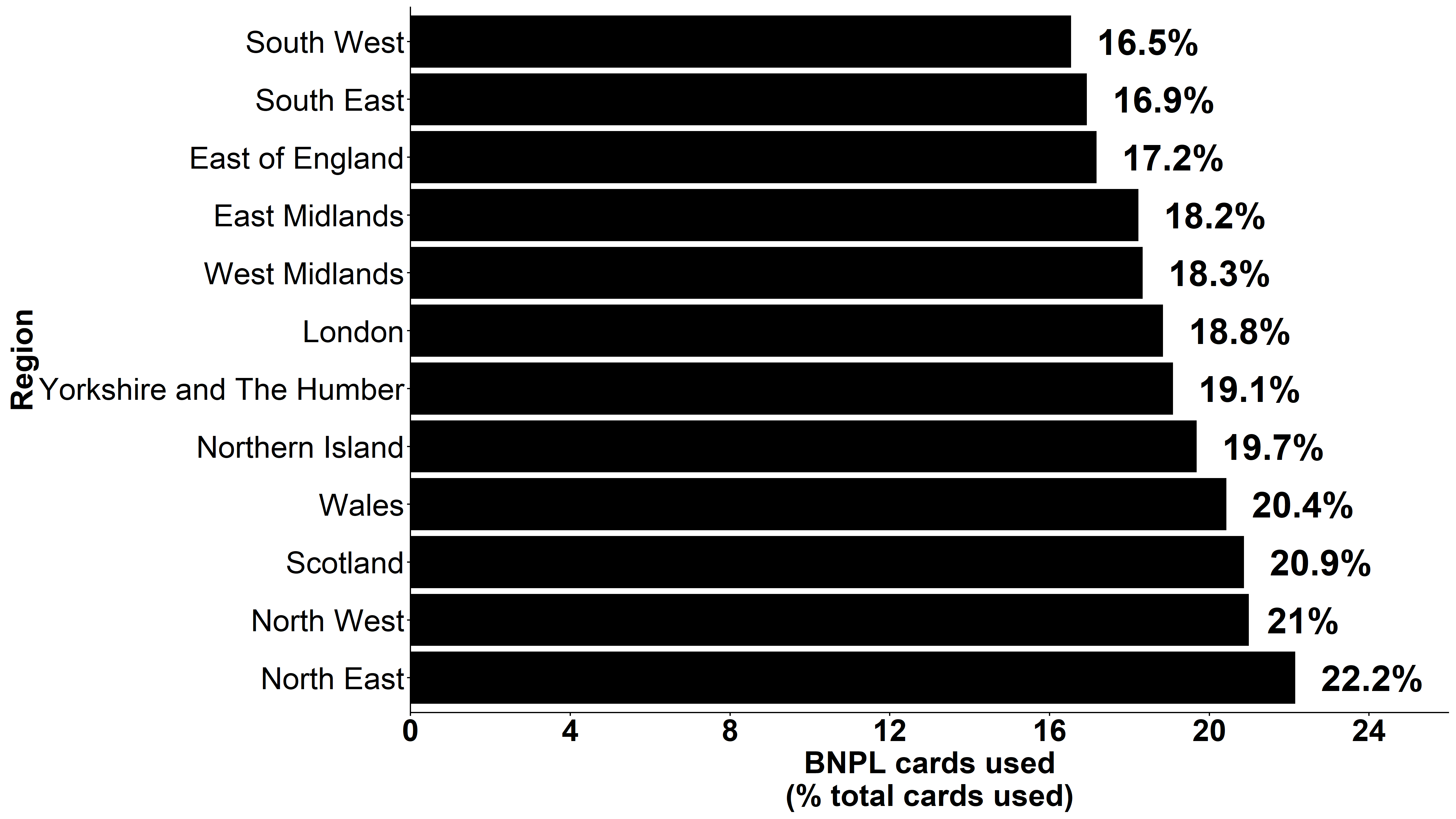}} \\ 
\textbf{B. Urban-rural area classification} \\ 
{\includegraphics[width=3.5in]{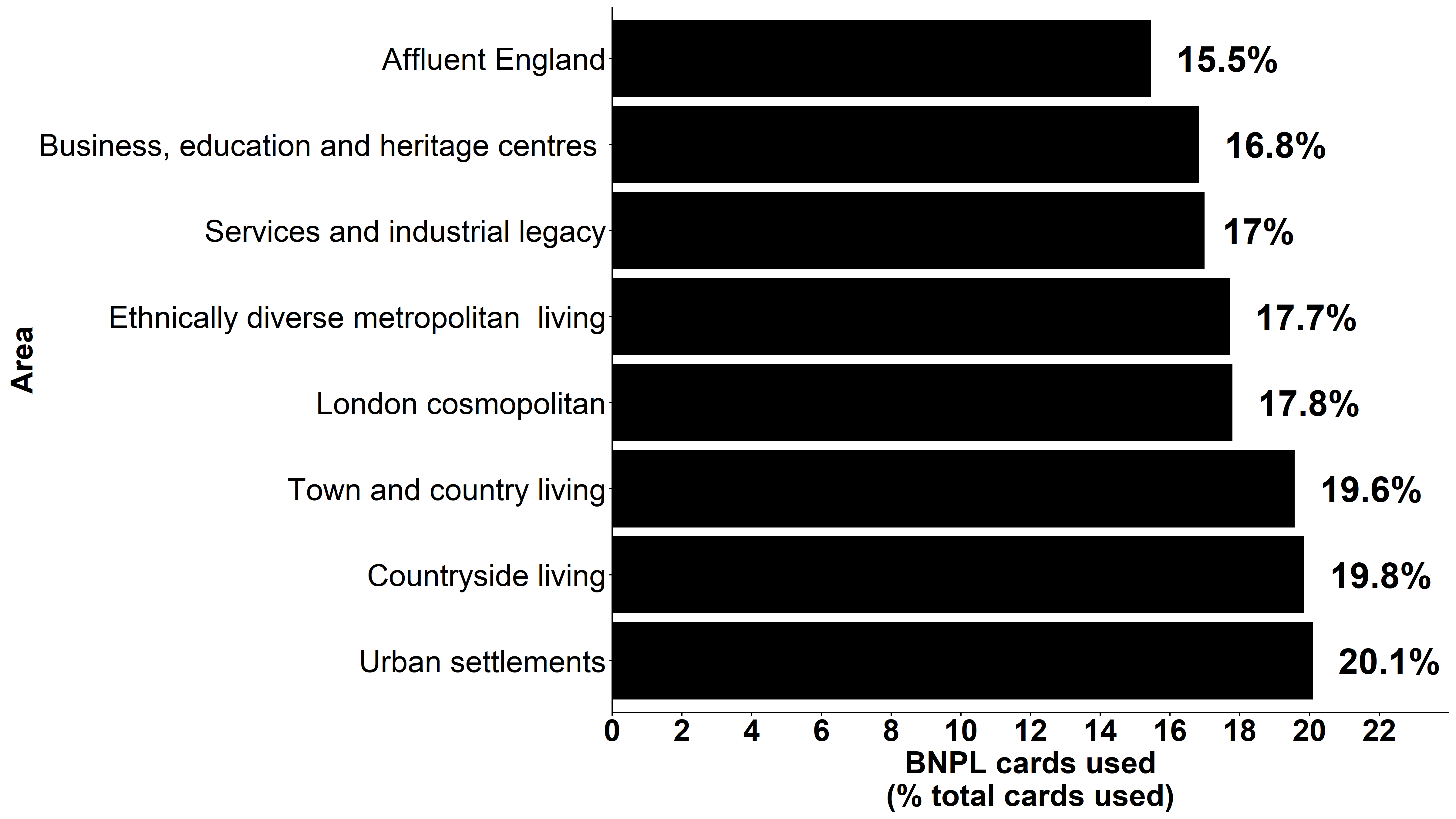}} \\ 
\end{tabular}
\end{center}

\begin{singlespace}
\noindent {\footnotesize \textit{Notes: UK credit card transactions and Office for National Statistics (ONS) data. BNPL is buy now, pay later. Actively-used credit cards defined as any BNPL or non-BNPL transactions in both January and December 2021. Panels allocate cards based on cardholder postcode sector to use ONS NUTS1 regions (Panel A) and ONS supergroup (2011) area classifications  (Panel B). Maps in Figure A8.}}
\end{singlespace}

\clearpage

\noindent \textbf{Figure A9. Maps of UK regions (A) and area classifications (B)} \\ 

\begin{center}
\begin{tabular}{cc} 
\textbf{A. Regions} & \textbf{B. Urban-rural area classification} \\ 
{\includegraphics[width=3in]{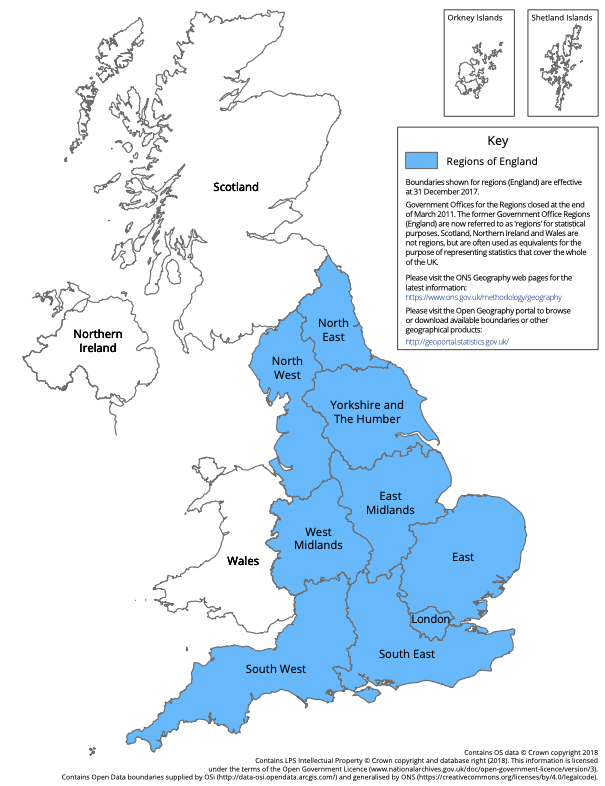}} & {\includegraphics[width=3in]{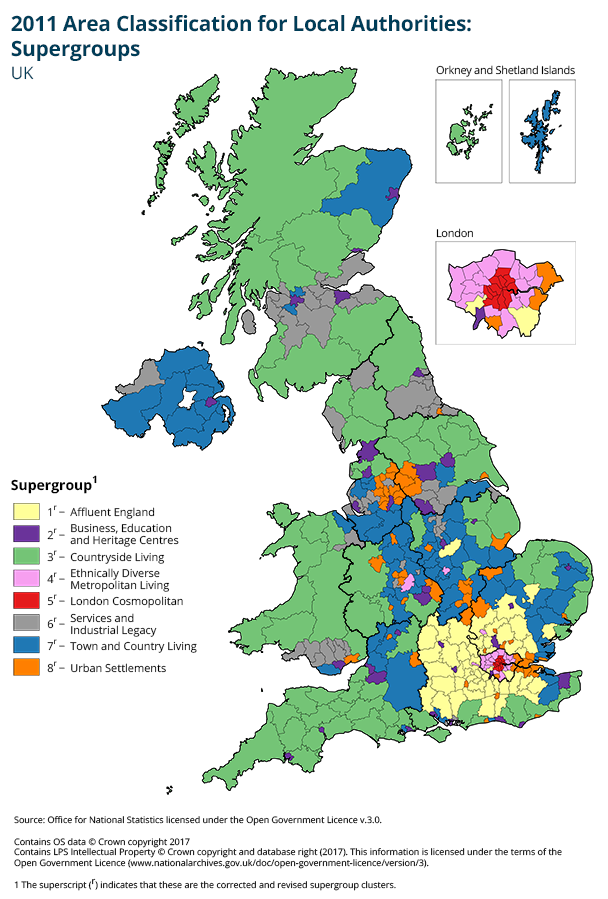}} \\
\end{tabular}
\end{center}

\begin{singlespace}
\noindent {\footnotesize \textit{Notes: Office for National Statistics (ONS) data. Definitions of ONS area classifications: \\  www.ons.gov.uk/methodology/geography/geographicalproducts/areaclassifications/2011areaclassifications}}
\end{singlespace}

\clearpage

\noindent \textbf{Figure A10. Distribution of BNPL spending for individual BNPL transactions (A) and aggregated across BNPL transactions (B) on active credit cards} \\ 

\begin{center}
\begin{tabular}{c} 
\textbf{A. Individual BNPL transaction values in 2021} \\ 
{\includegraphics[width=5.25in]{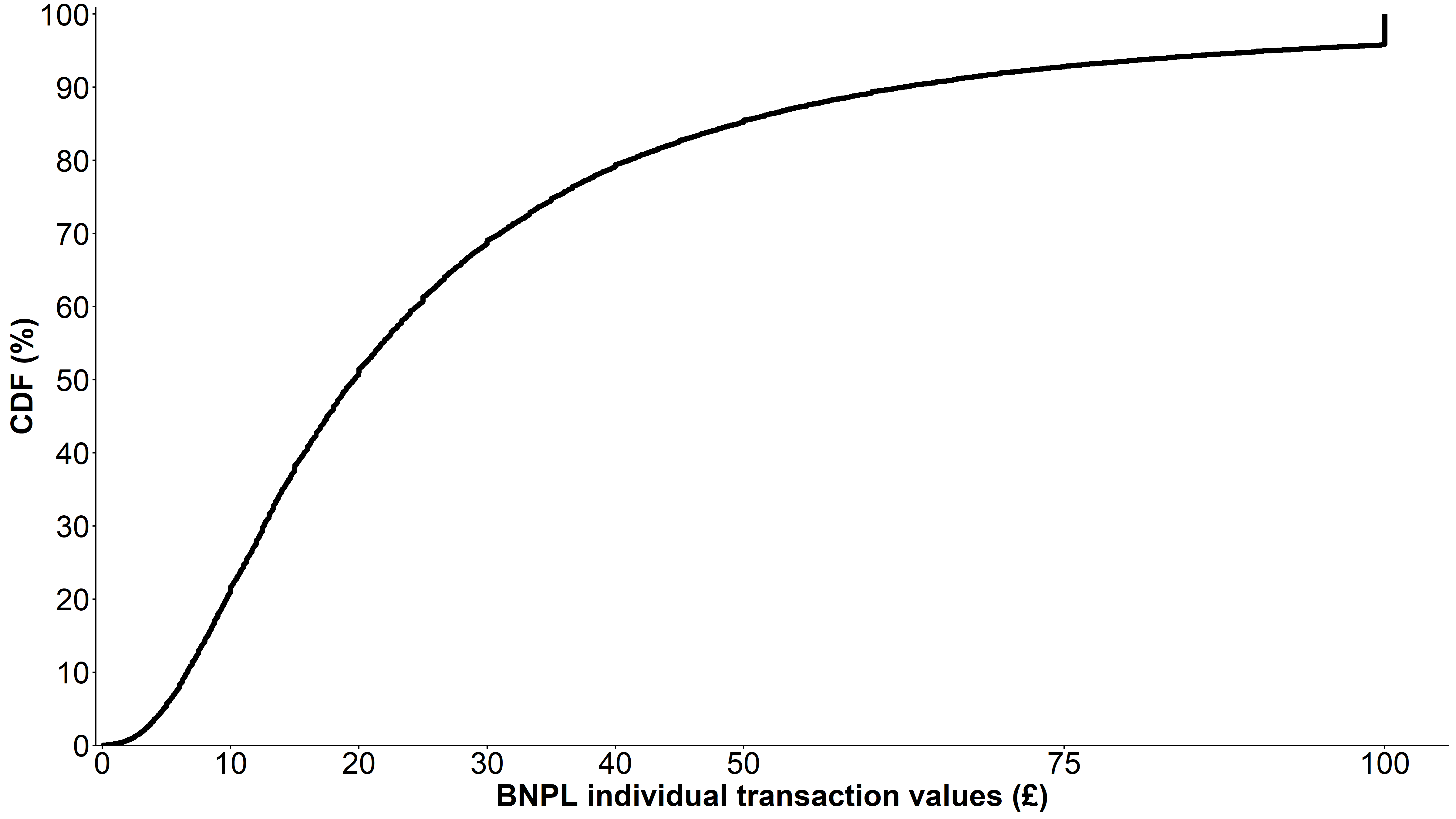}} \\
\textbf{B. Aggregated value of BNPL card-level spending in 2021} \\ 
{\includegraphics[width=5.25in]{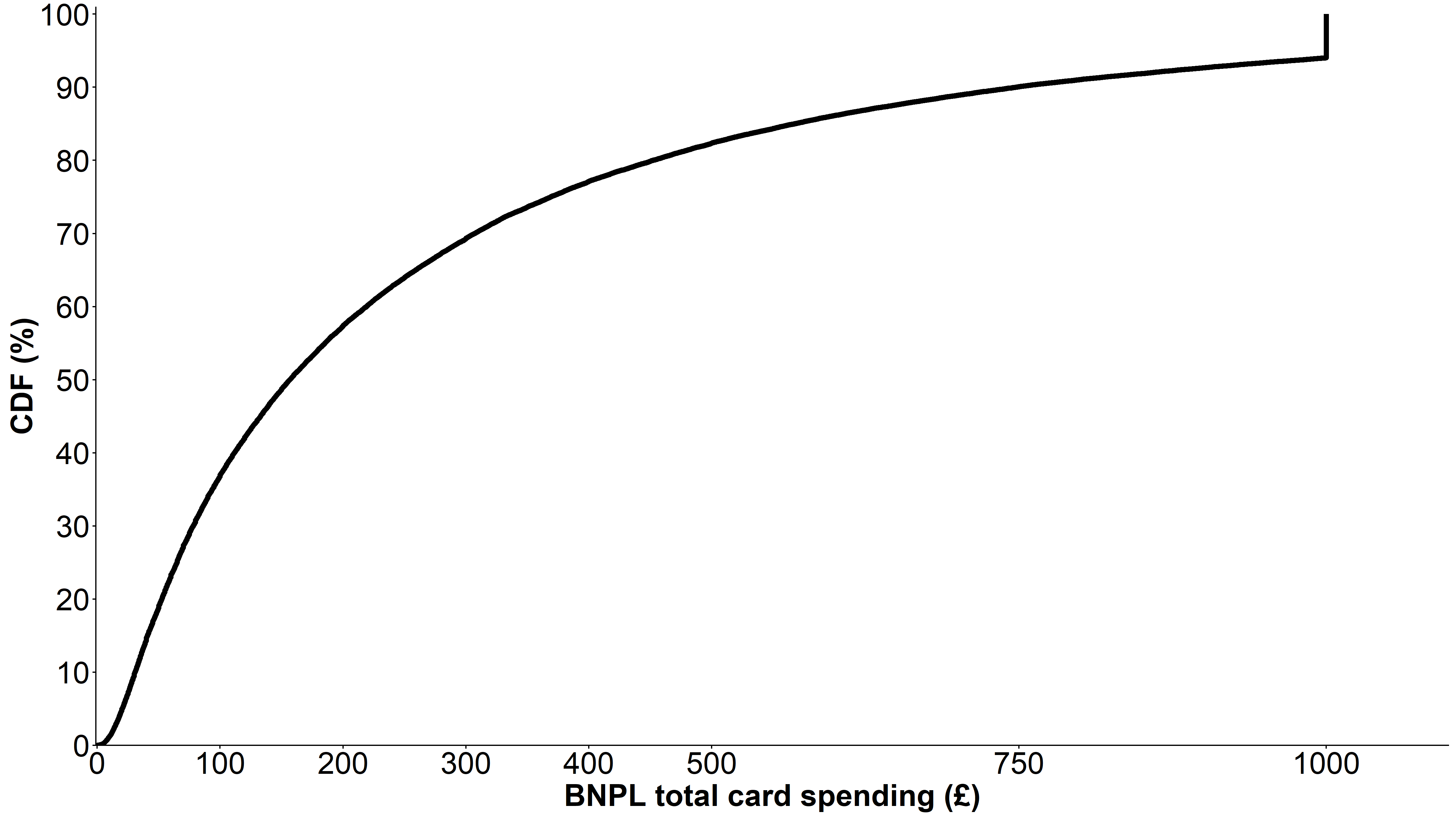}} \\
\end{tabular}
\end{center}

\begin{singlespace}
\noindent {\footnotesize \textit{Notes: UK credit card transactions data. Panel A and Panel B top-coded at \pounds100 and \pounds1,000 respectively. BNPL is buy now, pay later. Panels A and B include all BNPL transactions on credit cards that were active: defined as any BNPL or non-BNPL transactions in December 2021.}}
\end{singlespace}

\vspace{1cm}

\clearpage

\noindent \textbf{Figure A11. Robustness of distribution of BNPL spending for individual BNPL transactions (A) and aggregated across BNPL transactions on credit card (B) when active cards defined as transactions in both January and December 2021} \\ 

\begin{center}
\begin{tabular}{c} 
\textbf{A. Individual BNPL transaction values in 2021} \\ 
{\includegraphics[width=5.25in]{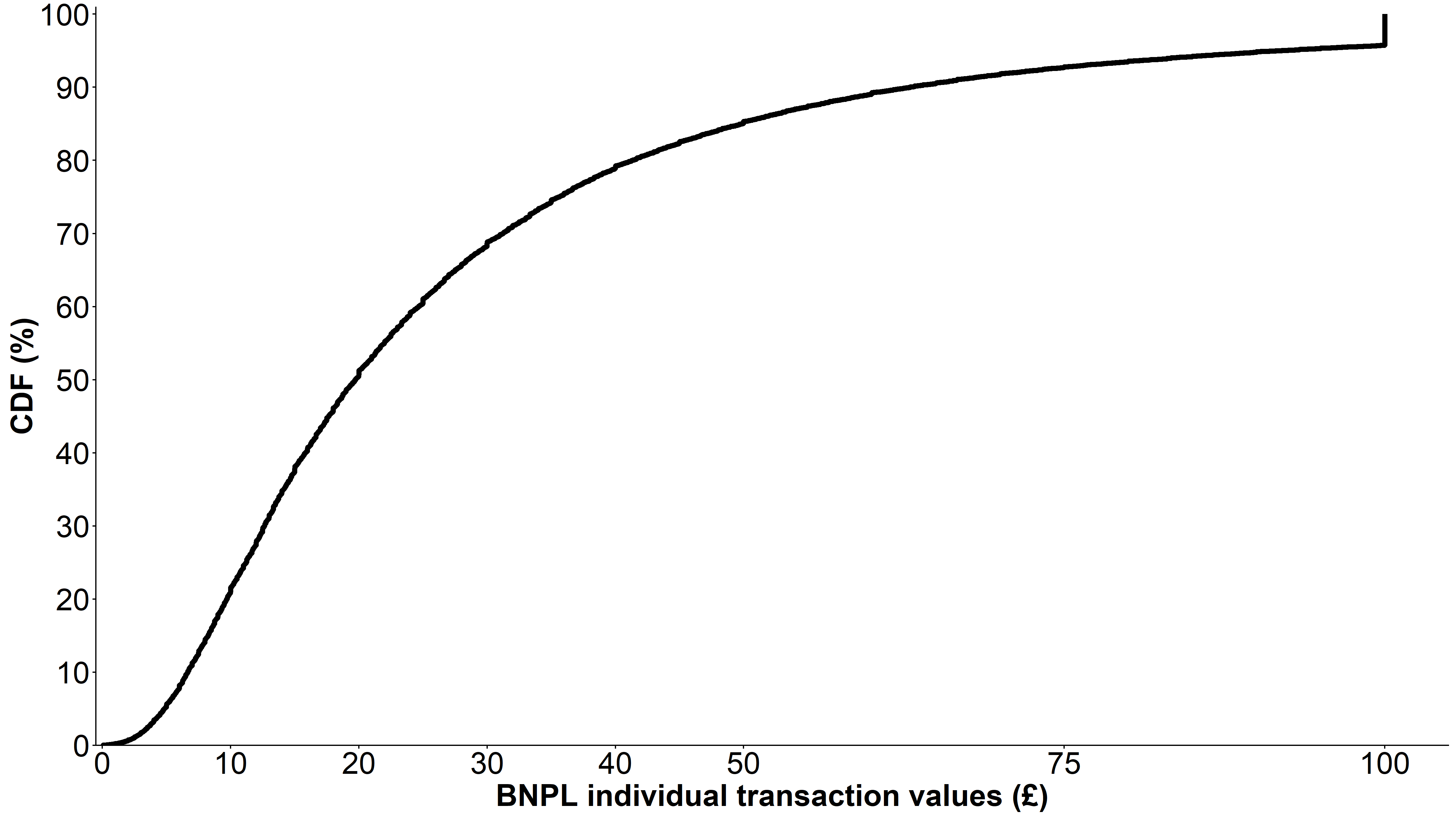}} \\
\textbf{B. Aggregated value of BNPL card-level spending in 2021} \\ 
{\includegraphics[width=5.25in]{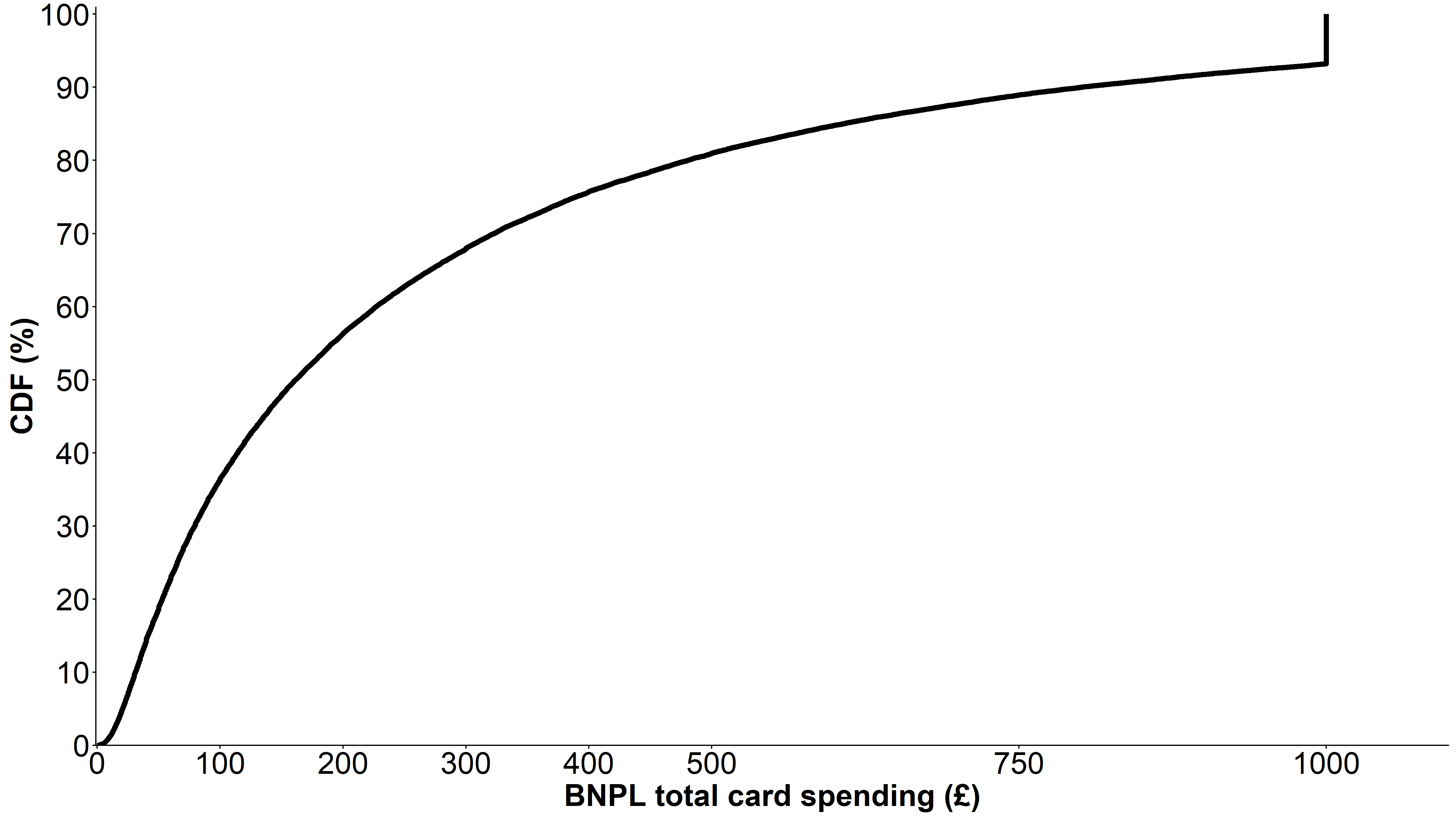}} \\
\end{tabular}
\end{center}

\begin{singlespace}
\noindent {\footnotesize \textit{Notes: UK credit card transactions data. Panel A and Panel B top-coded at \pounds100 and \pounds1,000 respectively. BNPL is buy now, pay later. Panels A and B include all BNPL transactions on credit cards that were active: defined as any BNPL or non-BNPL transactions in both January and December 2021.}}
\end{singlespace}

\clearpage
\bibliography{refs.bib}
\bibliographystyle{apalike}

\end{document}